\pdfoutput=1
\documentclass{article}

     \PassOptionsToPackage{numbers, compress}{natbib}

\usepackage[preprint]{neurips_2025}

\usepackage[utf8]{inputenc} %
\usepackage[T1]{fontenc}    %
\usepackage{hyperref}       %
\usepackage{url}            %
\usepackage{booktabs}       %
\usepackage{amsfonts}       %
\usepackage{nicefrac}       %
\usepackage{microtype}      %
\usepackage{xcolor}         %
\usepackage{graphicx}
\usepackage{wrapfig}

\usepackage{natbib} 

\usepackage{listings}
\usepackage{xcolor}
\usepackage{multirow}

\usepackage{algorithm}
\usepackage{algpseudocode}
\definecolor{comment_green}{HTML}{2CA02C}
\usepackage{amsmath}
\usepackage[capitalize,noabbrev]{cleveref}

\usepackage{tcolorbox}

\usepackage{xparse}
\usepackage{mdframed}
\usepackage{calc}

\definecolor{set2red}{HTML}{FC8D62}
\definecolor{set2green}{HTML}{66C2A5}
\definecolor{set2blue}{HTML}{8DA0CB}
\definecolor{fontc}{HTML}{403E30}
\definecolor{chatbg}{HTML}{F5F5F5}
\definecolor{chatsideline}{HTML}{20589F}

\newlength{\chatlogwidth}
\newlength{\chattextwidth}
\newlength{\chatinnerwidth}

\newcommand{\setspeakercolor}[1]{
  \ifnum#1=1
    \def\currentChatter{set2green}
  \else\ifnum#1=2
    \def\currentChatter{set2blue}
  \else\ifnum#1=3
    \def\currentChatter{set2red}
  \else
    \def\currentChatter{set2green}
  \fi\fi\fi
}

\newcommand{\setChatlogWidth}[1]{
  \setlength{\chatlogwidth}{#1}
  \setlength{\chatinnerwidth}{\chatlogwidth-35pt}
  \setlength{\chattextwidth}{0.90\chatinnerwidth}
}

\setChatlogWidth{\textwidth}

\newenvironment{chatlog}{
  \noindent
  \begin{minipage}{\chatlogwidth}
  \begin{mdframed}[
    backgroundcolor=chatbg,
    leftline=true,
    rightline=false,
    topline=false,
    bottomline=false,
    linecolor=chatsideline,
    linewidth=3pt,
    innerleftmargin=15pt,
    innerrightmargin=20pt,
    innertopmargin=10pt,
    innerbottommargin=5pt
  ]
}{
  \vspace{-0.2em}
  \end{mdframed}
  \end{minipage}
}

\NewDocumentCommand{\chatmsg}{O{1} m m}{
  \setspeakercolor{#1}
  \noindent{\textbf{\textcolor{\currentChatter!80!black}{#2}}}\par
  \vspace{0.35em}
  \noindent\hspace{1em}\parbox[t]{\chattextwidth}{\textcolor{fontc}{#3}}\par
  \vspace{0.7em}
}

\newcommand{\benchmark}{\textsc{AgentsNet}}

\usepackage[symbol]{footmisc}

\title{\benchmark{}: Coordination and Collaborative Reasoning in Multi-Agent LLMs}

\author{%
  Florian Grötschla$^*$$^\dagger$ \\
  ETH Zurich\\
  \And
  Luis Müller$^*$ \\
  RWTH Aachen University\\
  \And
  Jan Tönshoff$^*$ \\
  RWTH Aachen University\\
  \AND
  Mikhail Galkin \\
  Google Research\\
  \And
  Bryan Perozzi \\
  Google Research\\
}

\begin{document}

\maketitle

\footnotetext[1]{Equal contribution.}
\footnotetext[2]{Correspondence to: \texttt{fgroetschla@ethz.ch}}
\begin{abstract}
Large-language models (LLMs) have demonstrated powerful problem-solving capabilities, in particular when organized in multi-agent systems. However, the advent of such systems also raises several questions on the ability of a complex network of agents to effectively self-organize and collaborate. While measuring performance on standard reasoning benchmarks indicates how well multi-agent systems can solve reasoning tasks, it is unclear whether these systems are able to leverage their topology effectively. Here, we propose \benchmark{}, a new benchmark for multi-agent reasoning. By drawing inspiration from classical problems in distributed systems and graph theory, \benchmark{} measures the ability of multi-agent systems to collaboratively form strategies for problem-solving, self-organization, and effective communication given a network topology. We evaluate a variety of baseline methods on \benchmark{} including
homogeneous networks of agents which first have to agree on basic protocols for organization and communication. 
We find that some frontier LLMs are already demonstrating strong performance for small networks but begin to fall off once the size of the network scales. 
While existing multi-agent benchmarks cover at most 2--5 agents, \benchmark{} is practically unlimited in size and can scale with new generations of LLMs.
As such, we also probe frontier models in a setup with up to 100 agents.
\end{abstract}

\section{Introduction} \label{sec:introduction}
Human societies thrive on collaboration, with language serving as the primary medium through which individuals coordinate and achieve collective goals. From small teams to large-scale organizations, effective communication enables structured decision-making, problem-solving, and the emergence of complex behaviors that surpass the capabilities of any single individual. This interplay between communication and coordination is mirrored in computing, where distributed systems rely on structured information exchange to tackle problems that exceed the capacity of any single processor. Just as psychology studies individual cognition while sociology examines emergent behaviors in groups, distributed systems research focuses on multi-agent coordination beyond what a single machine can accomplish~\citep{lenzen2012distributed}.

Recently, distributed systems have been playing an increasingly important role in AI through the emergence of general-purpose multi-agent systems built on top of large language and vision models (LLMs).
Agent-based frameworks such as generative agents \citep{park2023generative} have demonstrated the potential of solving complex problems with LLM-based agents. In particular, it has been shown that networks of LLM-based agents can outperform single agents \citep{chen2024agentverse, qian2024scaling, zhuge2024gptswarm, marro2024scalable}, mirroring aspects of human teamwork. For example, GPTSwarm \citep{zhuge2024gptswarm} introduce a graph-based approach inspired by language-based societies of mind \citep{zhuge2023mindstorms}, demonstrating that organizing LLM-based agents in structured topologies enhances their performance on benchmarks like MMLU~\citep{hendrycks_mmlu}, HumanEval~\citep{chen2021codex}, and GAIA~\citep{mialon2024gaia}. Yet despite promising results from structured agent networks, existing benchmarks fall short in evaluating the core competencies of multi-agent systems: scalable coordination, decentralized communication, and collaborative reasoning. To address this gap, we introduce \benchmark{}, a principled multi-agent benchmark that measures these capabilities across diverse network structures and scales.

\begin{wrapfigure}{r}{0.55\textwidth}
    \centering
    \includegraphics[width=1.0\linewidth]{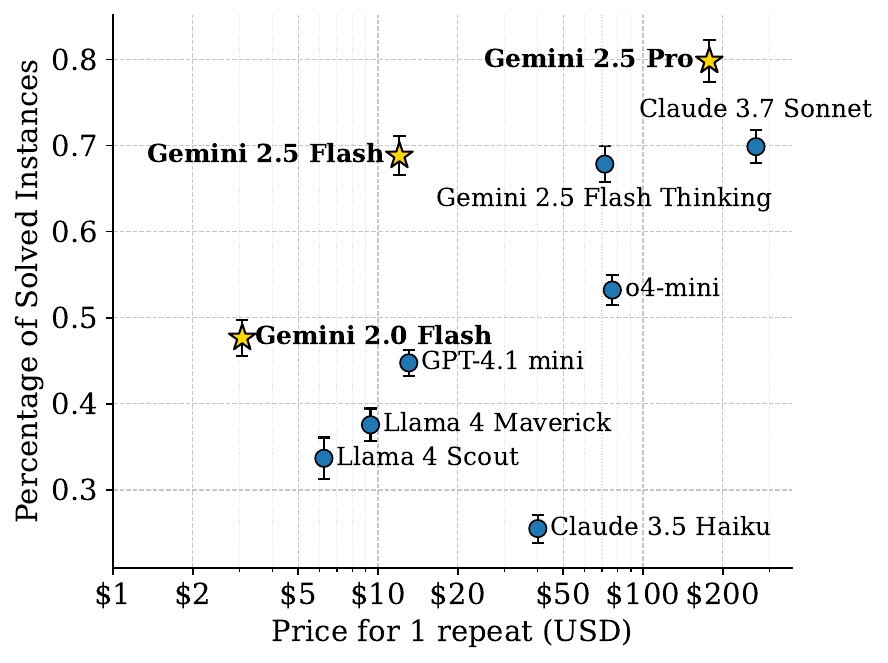}
    \caption{Mean \benchmark{} score of models versus API costs per repeat (May 15, 2025). Error bars indicate standard error of the mean. Gold stars denote Pareto-optimal models.%
    }
    \label{fig:performance_by_cost}
\end{wrapfigure}

\benchmark{} assesses the agent's coordinative and collaborative capabilities through fundamental problems in distributed computing. Concretely, we identify five central problems from the distributed systems literature to construct corresponding coordination and collaboration tasks for multi-agent systems. Solving these tasks requires anything from local information aggregation to global coordination over multiple communication rounds.
As a canonical example, whenever multi-agent systems are tasked with solving a certain problem, agents must necessarily be able to reach an agreement on the solution, a problem known in fault-tolerant distributed computing as consensus \citep{fischer1985impossibility}. 
In another example, agents first agree on a single agent to take a leadership role, and then subsequently solve the task, guided and instructed by the elected leader. Selecting a single leader in a network is known as the leader election problem \citep{angluin1980local}.
Fortunately, such problems are well-studied and theoretically grounded, providing an ideal testbed for the coordination and collaboration skills of multi-agent systems.

Various multi-agent benchmarks exist, but no benchmark explicitly assesses the ability of multi-agent systems for structured coordination and collaboration, which should be seen as fundamental capabilities of effective distributed systems. As such, \benchmark{} complements the existing suite of multi-agent benchmarks of LLMs with a particular focus on grounding in distributed systems theory, network topology, and scalability to large agent networks. Concretely, we make the following contributions:
\begin{enumerate}
    \item We build \benchmark{} from \textit{graph coloring, minimal vertex cover, maximal matching, leader election}  and \textit{consensus}: five fundamental distributed computing problems that evaluate the ability of multi-agent systems to effectively self-organize, coordinate, and communicate to solve complex reasoning problems.
    \item We design a robust and scalable message-passing protocol for effective agent-to-agent communication and evaluate on a rich set of graph instances, sampled from various graph models such as small-world \citep{watts1998collective} or preferential attachment models \citep{barabasi1999emergence}, which capture structural properties of real-world networks.
    \item We evaluate a variety of agentic baselines on \benchmark{}, ranging from open-source LLMs such as Llama 4, to frontier models such as GPT, Gemini, and Claude, as well as the latest reasoning models, on the graphs of 4, 8, 16 nodes scaling the problem size to 100 agents which is well beyond existing agentic benchmarks.
    \item We provide an in-depth qualitative analysis and highlight the challenges in coordinative and collaborative capabilities of LLMs 
    to further improve multi-agent systems.
\end{enumerate}

\section{Related Work}
Ensembling multiple agents to collaboratively negotiate solutions has emerged as an effective paradigm to improve LLM performance on complex tasks~\citep{du2023improving,xiong2023examining,liang2024encouraging}.
This has been extended through work on different network topologies for more structured agent interaction. 
Some studies examine pre-determined graph structures \citep{regan2024problem,qian2024scaling} while others propose automatically adapting network topology \citep{liu2023dynamic,chen2024agentverse,zhuge2024gptswarm}. 
Experiments show different topologies perform best for specific tasks \citep{chen2024agentverse,zhuge2024gptswarm} and large-scale LLM agent networks exhibiting known social phenomena \citep{yang2024oasisopenagentsocial,chuang2024simulating}.
Parallel research examines LLMs' ability to reason with graph-structured data. 
Studies propose evaluation datasets \citep{fatemi2024talk,wang2024can,zhang2024llm4dyg,tang2025evaluating,skianis2024graphreasoninglargelanguage} using single-agent setups where graphs are encoded as text. \citet{fatemi2024talk} investigate graph encoding methods, \citet{sanford2024understanding} categorize graph reasoning problems by complexity, while \citet{wang2024can} and \citet{skianis2024graphreasoninglargelanguage} explore effective prompting techniques.
Our work bridges these research directions by studying multi-agent systems solving graph reasoning problems collaboratively.
Our benchmark is complementary to recent agentic benchmarks~\citep{liu2024agentbench,yin2024mmau,agashe2024llm,yao2024tau,ni2025coral} but scales to a practically unlimited number of agents due to the generative problem creation protocol, with experiments involving up to 100 coordinating agents.
Related human studies on decentralized problem-solving in social networks show that network topology and size strongly influence coordination success \citep{kearns2006experimental,judd2010behavioral,chiang2024adaptive}.
\Cref{appendix:extended_rw} provides an extended discussion of related studies.

\begin{figure}
    \centering
    \includegraphics[width=1.0\linewidth]{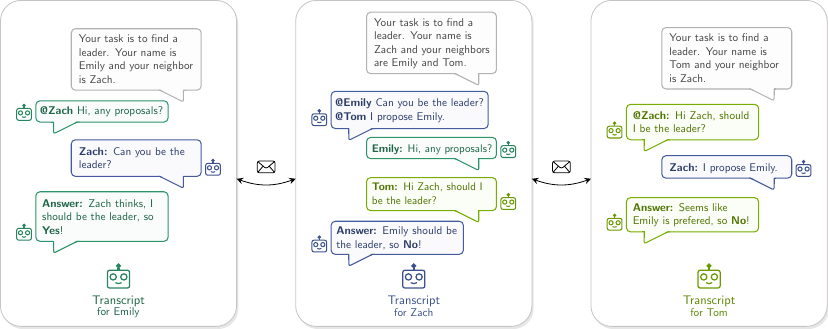}
    \caption{Example communication between three agents on a simplified topology. Agents Emily, Zach, and Tom each receive and send messages to their neighbors in multiple rounds of message-passing; see \Cref{app:qualitative} for an in-depth qualitative analysis of transcripts.}
    \label{fig:robochat}
\end{figure}

\section{Tasks, Evaluation, and Graph Models}\label{sec:tasks_graphs_evaluation}
To evaluate the ability of multi-agent systems to self-organize, coordinate, and communicate effectively, we design a benchmark consisting of fundamental problems from distributed computing. These problems span a range of complexities, from local tasks that require minimal coordination to global problems that necessitate multi-round communication. In what follows, we introduce the theoretical problems and describe how we map each problem to a corresponding agentic task. Afterwards, we introduce the graph distributions used within \benchmark{}.

\subsection{Benchmarking Tasks}
We evaluate multi-agent systems on a set of distributed computing problems that test their ability to aggregate information, self-organize, and coordinate. These tasks are selected for their foundational nature in distributed computing and for their relevance as naturally appearing subproblems in more complicated tasks. They span a diverse range of coordination requirements and communication complexities, from purely local information exchange to global decision-making; see \Cref{tab:theory_problem_overview} for an overview of the different theoretical problems selected for \benchmark{}. %

\begin{figure}
    \centering
    \includegraphics[width=\linewidth]{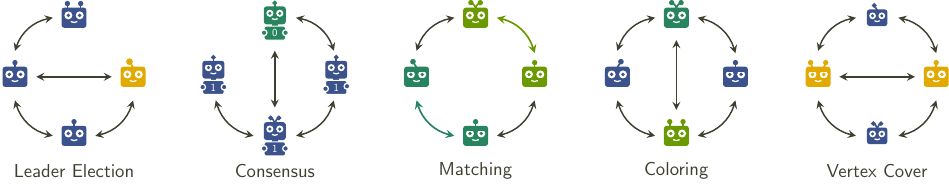}
    \caption{Overview of the tasks in \benchmark{}: In \textsc{LeaderElection}, the task is to select a single agent as the leader of the network. In \textsc{Consensus}, the task is for all agents to agree on a specific value, for example $0$ or $1$. In \textsc{Matching}, the task is for pairs of agents to team up without conflicts. In \textsc{Coloring}, the task is for agents to select a group (indicated by a color), such that none of their neighbors are in the same group as them. In \textsc{VertexCover}, the task is to find a minimal group of coordinator agents such that each agent is a neighbor to at least one coordinator.}
    \label{fig:task_overview}
\end{figure}

\paragraph{($\Delta + 1$)-Coloring.} Each node is assigned a color using at most $\Delta + 1$ colors, where $\Delta$ is the maximum node degree. This problem has a well-defined distributed complexity of $O(\log^* n)$ in bounded-degree graphs~\citep{barenboim2016deterministic}.
This task is particularly useful for role assignment within multi-agent systems. For instance, agents can be designated to perform specific sub-tasks (e.g., web search, reasoning, coding, planning), with the constraint that directly connected agents are assigned distinct roles to avoid redundancy. Solving this task reflects the system's ability to efficiently distribute responsibilities across the network with minimal overlap in capabilities. The corresponding agentic task is to form groups, with a pre-defined number of groups, and where each group corresponds to a color. After message-passing, each agent chooses the group it wants to be in. The task is solved if the groups form a valid $\Delta + 1$-coloring. In \benchmark{}, we refer to this task as \textsc{Coloring}.

\paragraph{Minimal Vertex Cover.} A minimal vertex cover is a subset of nodes such that every edge in the graph has at least one endpoint in the subset, and removing any node from this subset would violate that property. This problem has a close relationship with the maximal independent set and is similarly fundamental in distributed computing, with known randomized solutions in $O(\log^* n)$ rounds~\cite{linial1992local}.
In agentic networks, a minimal vertex cover can represent a minimal set of monitor or gateway agents that maintain awareness of all interactions in the system. These agents could take on responsibilities such as relaying messages, auditing behavior, or bridging subgroups. The task tests the ability to identify a compact yet effective set of nodes with high influence or observability. The corresponding agentic task is to select a group of coordinators among the agents. After message-passing, each agent is asked whether it is a coordinator. The agents can respond with either \textit{Yes} or \textit{No}. The task is solved if coordinators form a minimal vertex cover. In \benchmark{}, we refer to this task as \textsc{VertexCover}.

\paragraph{Maximal Matching.} A maximal matching is a set of edges such that no two edges share a vertex, and no additional edges can be added without violating this property. This task captures the ability of agents to negotiate pairwise agreements without global knowledge, which is useful in scenarios where resource allocation or mutual exclusivity must be enforced (e.g., agent-to-agent task assignment).
Randomized algorithms typically solve this problem in $O(\log^* n)$ rounds~\cite{peleg2000distributed}.  The corresponding agentic task is for the agents to form pairs. After message-passing, each agent is asked to name the neighbor it wants to pair up with. The agents can also respond with \textit{None} if they cannot find a match (all neighbor agents are already paired up with other agents). The task is solved if the paired agents form a maximal matching.
In \benchmark{}, we refer to this task as \textsc{Matching}.

\paragraph{Leader Election.} One node must be selected as the leader, while all others acknowledge that they are not. This classic coordination task is central to evaluating how well agents establish hierarchy and delegate global decision-making~\citep{angluin1980local}.
In multi-agent systems, leader election can be interpreted as selecting a central planner or controller agent responsible for strategy synthesis, while the remaining agents act as executors. Effective leader election demonstrates the system's capacity to break symmetry and converge on a single authority. In general graphs, the round complexity is $O(D)$, where $D$ is the network diameter~\cite{lynch1996distributed}. The corresponding agentic task is to select a single leader among the agents. After message-passing, each agent is asked whether it is the leader. The agents can respond with either \textit{Yes} or \textit{No}. The task is solved if there exists exactly one leader.
In \benchmark{}, we refer to this task as \textsc{LeaderElection}.

\begin{wraptable}{r}{0.5\textwidth}
    \centering
    \begin{tabular}{cc} \toprule
       Graph Problem  & Round Complexity \\ \midrule
        $(\Delta + 1)$-Coloring & $\Omega(\log^*(n))$\\
        Minimal Vertex Cover & $\Omega(\log^*(n))$\\
        Leader election & $\Omega(D)$\\
        Maximal Matching & $\Omega(\log^*(n))$\\
        Consensus & $\Omega(D)$\\
        \bottomrule \\
    \end{tabular}
    \caption{Overview of the theoretical problems from distributed computing that form the basis of \benchmark{}, together with (not necessarily tight) theoretical lower bounds for their round complexity in the randomized LOCAL~\citep{linial1992local} model.}
    \label{tab:theory_problem_overview}
\end{wraptable}

\paragraph{Consensus.}
In the consensus problem, all agents must agree on a single value from the set $\{0, 1\}$. In our benchmark, we focus on the basic setting without any faulty or Byzantine agents. The goal is for all agents to coordinate and produce the same final answer after a number of communication rounds. A successful solution requires that every agent outputs the same value, either 0 or 1. This task tests the ability of multi-agent systems to converge to a global agreement through local message-passing alone. In synchronous networks, achieving consensus generally requires $\mathcal{O}(D)$ rounds~\cite{lynch1996distributed}. The corresponding agentic task is to choose between a value $0$ and $1$. After message-passing, each agent is asked to announce its selected value. The task is solved if all agents announce the same value. In \benchmark{}, we refer to this task as \textsc{Consensus}.

Together, these tasks cover a broad spectrum of problems known in the distributed 
computing literature, which allows \benchmark{} to evaluate the reasoning, communication, and organizational capabilities of multi-agent systems.

\subsection{Network Topologies}
While classical distributed computing often studies problems on random graphs such as Erd\H{o}s-Renyi networks~\citep{erdos1960evolution}, these do not adequately capture the structural properties of real-world networks. Instead, we focus on three well-established graph models, namely the Watts-Strogatz graphs~\citep{watts1998collective} (\textsc{SmallWorld}) exhibiting both short average path lengths and high clustering coefficients; preferential attachment graphs~\citep{barabasi1999emergence} (\textsc{ScaleFree}) containing hubs (high-degree nodes) and follow a power-law degree distribution; geometric graphs by constructing a Delaunay triangulation over randomly sampled 2D points, (\textsc{Delaunay}), maintaining a spatial relationship between nearby agents. We describe these graph models in more detail in \Cref{app:graph_models}.

\section{Agent-to-Agent Communication via Message-Passing}\label{sec:mp_protocol}
To systematically study how agents exchange information and collaborate, we employ a communication model that draws inspiration from classical distributed computing, while adapting to the capabilities and constraints of modern LLM-based agents.
Our setup is based on the LOCAL model~\citep{linial1992local} from distributed algorithms, in which the computation proceeds in synchronous rounds and each agent can exchange messages only with its immediate neighbors on the communication graph. Agents must base their decisions exclusively on local information aggregated over multiple rounds of interaction. This model captures fundamental aspects of decentralized reasoning, where global strategies emerge from purely local exchanges without centralized control. 
Unlike nodes in deterministic systems, LLM-based agents exhibit stochastic behavior due to inherent randomness in their generation processes. This means that our model is most closely aligned with the randomized version of the LOCAL model.

Given a communication network, each node, that is, each agent, is instantiated as an instruction-tuned LLM that interfaces with its neighbors through a structured chat history. Initially, we provide each agent with a \textit{system prompt} detailing the task, for example, \textsc{Coloring}, the rules of message-passing, the names of its neighbors, and a notification that the agent must output a result in its \textit{final response} after a fixed number of rounds of message-passing; see \Cref{app:implementation} for the full system prompt.
We describe task descriptions, message-passing rules, and final response crafting below.

\paragraph{Task description.}
For each task, we provide a short description of the task, as well as which information we seek to extract in the final response. For example, for \textsc{LeaderElection}, we provide the following task description:

\setChatlogWidth{\textwidth}
\begin{chatlog}
\chatmsg[1]{System}{\texttt{\small Your task is to collaboratively solve the problem of electing a single leader. [...] You will be requested to state whether or not you are the leader. The response should either be 'Yes' or 'No'. The final result should be such that exactly one agent responds with 'Yes' and all others say 'No' as there should be exactly one leader.}}
\end{chatlog}

Note that the "[...]" indicates that different parts of the task description appear at different parts of the system prompt.

\paragraph{Message-Passing Rules.}
For message-passing, we iteratively prompt each agent with the current chat history, including the latest messages received from its neighbors, to generate new messages to each neighbor in the form of a flat JSON. Here, each key corresponds to the name of a neighboring agent, and each value to the message intended for the corresponding neighbor. Optionally, we also ask the model to elaborate its chain-of-thought before responding.
An example of this message exchange can look as follows:

\begin{chatlog}
\chatmsg[3]{Human}{\texttt{\small These are the messages from your neighbors: Message from Emma: Hello Evelyn, this is Emma. I appreciate your response and [...] Message from Dorothy: [...] 
Elaborate your chain of thought step-by-step first, then output the messages for your neighbors. Output your messages in JSON format as specified earlier.}}
\end{chatlog}

In practice, and in particular for smaller models, we observe that agents sometimes fail to output valid JSON. In such cases, we simply ask the model to try again using the entire chat history, including the incorrect answer given by the model, as well as a prompt to retry.

\paragraph{Final Response.}
After a fixed number message-passing rounds, we ask the model to give its task-specific response based on the chat history accumulated during message-passing. Again, we ask models for a structured output, this time using a simpler, string-based format. For example, for \textsc{LeaderElection}, the final response prompt is:

\vspace{2px}
\begin{chatlog}
\chatmsg[3]{Human}{\texttt{\small Are you the leader? Format your answer as follows: '\#\#\# Final Answer \#\#\#', followed by your final answer. Don't use any text for your final answer except one of these valid options: 'Yes', 'No'.}}
\end{chatlog}

Once more, we find that models generate a valid response after at most one retry.
The benchmarking results are then computed from these final answers, following the task-specific evaluation methods described in \Cref{sec:tasks_graphs_evaluation}.

\section{Experiments}\label{sec:experiments}

\begin{table}
    \centering
    \resizebox{\linewidth}{!}{
    \begin{tabular}{lccccc|c}
    \toprule
       Model  & \scriptsize \textsc{Coloring} & \scriptsize \textsc{Consensus} & \scriptsize \textsc{Leader Election} & \scriptsize \textsc{Matching} & \scriptsize \textsc{Vertex Cover} & \scriptsize \benchmark{}  \\
       \midrule
Claude 3.5 Haiku & 0.14 \tiny \textcolor{gray}{(0.04)} & 0.69 \tiny \textcolor{gray}{(0.05)} & 0.19 \tiny \textcolor{gray}{(0.03)} & 0.18 \tiny \textcolor{gray}{(0.03)} & 0.08 \tiny \textcolor{gray}{(0.03)} & 0.26 \tiny \textcolor{gray}{(0.02)} \\
Claude 3.7 Sonnet & 0.58 \tiny \textcolor{gray}{(0.05)} & 1.00 \tiny \textcolor{gray}{(0.00)} & 0.96 \tiny \textcolor{gray}{(0.03)} & 0.55 \tiny \textcolor{gray}{(0.06)} & 0.40 \tiny \textcolor{gray}{(0.05)} & 0.70 \tiny \textcolor{gray}{(0.02)} \\
GPT-4.1 mini & 0.05 \tiny \textcolor{gray}{(0.02)} & 0.99 \tiny \textcolor{gray}{(0.01)} & 0.86 \tiny \textcolor{gray}{(0.05)} & 0.12 \tiny \textcolor{gray}{(0.03)} & 0.22 \tiny \textcolor{gray}{(0.04)} & 0.45 \tiny \textcolor{gray}{(0.01)} \\
Gemini 2.0 Flash & 0.32 \tiny \textcolor{gray}{(0.05)} & 0.85 \tiny \textcolor{gray}{(0.04)} & 0.69 \tiny \textcolor{gray}{(0.05)} & 0.36 \tiny \textcolor{gray}{(0.05)} & 0.16 \tiny \textcolor{gray}{(0.04)} & 0.48 \tiny \textcolor{gray}{(0.02)} \\
Gemini 2.5 Flash & 0.39 \tiny \textcolor{gray}{(0.06)} & 1.00 \tiny \textcolor{gray}{(0.00)} & 1.00 \tiny \textcolor{gray}{(0.00)} & 0.55 \tiny \textcolor{gray}{(0.04)} & 0.50 \tiny \textcolor{gray}{(0.09)} & 0.69 \tiny \textcolor{gray}{(0.02)} \\
Gemini 2.5 FT & 0.53 \tiny \textcolor{gray}{(0.05)} & 0.99 \tiny \textcolor{gray}{(0.01)} & 0.98 \tiny \textcolor{gray}{(0.02)} & 0.47 \tiny \textcolor{gray}{(0.02)} & 0.43 \tiny \textcolor{gray}{(0.09)} & 0.68 \tiny \textcolor{gray}{(0.02)} \\
Gemini 2.5 Pro & 0.62 \tiny \textcolor{gray}{(0.07)} & 0.99 \tiny \textcolor{gray}{(0.01)} & 0.89 \tiny \textcolor{gray}{(0.06)} & 0.75 \tiny \textcolor{gray}{(0.05)} & 0.73 \tiny \textcolor{gray}{(0.06)} & 0.80 \tiny \textcolor{gray}{(0.02)} \\
Llama 4 Maverick & 0.20 \tiny \textcolor{gray}{(0.04)} & 0.85 \tiny \textcolor{gray}{(0.04)} & 0.56 \tiny \textcolor{gray}{(0.06)} & 0.20 \tiny \textcolor{gray}{(0.04)} & 0.07 \tiny \textcolor{gray}{(0.03)} & 0.38 \tiny \textcolor{gray}{(0.02)} \\
Llama 4 Scout & 0.21 \tiny \textcolor{gray}{(0.06)} & 0.67 \tiny \textcolor{gray}{(0.05)} & 0.38 \tiny \textcolor{gray}{(0.06)} & 0.30 \tiny \textcolor{gray}{(0.05)} & 0.13 \tiny \textcolor{gray}{(0.04)} & 0.34 \tiny \textcolor{gray}{(0.02)} \\
o4-mini & 0.22 \tiny \textcolor{gray}{(0.04)} & 0.92 \tiny \textcolor{gray}{(0.04)} & 0.92 \tiny \textcolor{gray}{(0.03)} & 0.33 \tiny \textcolor{gray}{(0.04)} & 0.27 \tiny \textcolor{gray}{(0.04)} & 0.53 \tiny \textcolor{gray}{(0.02)} \\
       \bottomrule
    \end{tabular}}
    \caption{Fraction of solved instances together with standard error over multiple i.i.d. samples from the same graph distribution (in gray) on \benchmark{}. Gemini 2.5 FT = Gemini 2.5 Flash Thinking.}
    \label{tab:main_results}
\end{table}

With the building blocks of \benchmark{} established in \Cref{sec:tasks_graphs_evaluation}, and our message-passing protocol described in \Cref{sec:mp_protocol}, we are now ready to describe our benchmark design and present our results.%

\subsection{Setup}
For benchmarking, we generate a set of 27 network topologies, consisting of 9 small-world, scale-free, and Delaunay graphs, respectively, ranging in size from 4 to 16 nodes. Concretely, for each graph size in $\{4, 8, 16\}$ and each graph distribution in $\{\textsc{SmallWorld}, \textsc{ScaleFree}, \textsc{Delaunay}\}$, we generate three graphs. Further, we determine the number of message-passing rounds as follows. For our global tasks, \textsc{LeaderElection} and \textsc{Consensus}, each agent must be able to exchange information with the entire network. Hence, for those two tasks, we select the number of message-passing rounds as $2D+1$, where $D$ is the diameter of the graph, to ensure that each pair of agents is able to exchange messages at least once. For the local tasks, \textsc{Coloring}, \textsc{Matching}, and \textsc{VertexCover}, we determine the number of rounds based on the graph size. Specifically, for graphs with 4 nodes, we choose 4 rounds, for 8 nodes -- 5 rounds, for  16 nodes -- 6 rounds.

\paragraph{Models.}
We evaluate a variety of frontier LLMs on \benchmark{}, including Claude 3.5 Haiku and Claude 3.7 Sonnet \citep{anthropic2024claude3}, 
Gemini 2.0 Flash \citep{google2024gemini2.0}, 
Gemini 2.5 Flash~\citep{google2024gemini2.5flash}, GPT-4.1-mini \citep{openai2025gpt4.1}, as well as Llama 4 Maverick and Scout \citep{meta2025llama4}, as representative open-source models. Notably, we include both large instruction-tuned models as well as reasoning models such as Gemini 2.5 Flash Thinking, Gemini 2.5 Pro \citep{google2024gemini2.5pro}, and o4-mini~\citep{openai2025o4}. The choice of models is motivated by an effective context window larger than 16K tokens, as problems on graphs of 8 and 16 nodes, especially at later stages of message passing, accumulate a long communication history.

\paragraph{Evaluation.}
\benchmark{} uses a binary evaluation metric, counting only fully correct solutions where the entire agent network satisfies the task specification. This strict criterion reflects the nature of distributed computing problems, where partial correctness often does not imply successful coordination. For example, in \textsc{Coloring}, most nodes may be correctly colored by chance, but only a valid global coloring confirms coordinated conflict resolution. However, in \Cref{app:benchmark_tasks}, we also discuss and report the soft evaluation scores to obtain a more continuous measure of the quality of responses, motivated by the findings in \citet{schaeffer2023emergent}, that emergent behaviors can often be explained by discontinuous metrics.

For each task and graph size, we sample three graphs per topology (small-world, scale-free, Delaunay) and run at least one repeat per graph. We report the mean of solved runs and the standard errors of the mean, computed across these runs~\cite{miller2024adding}. Details on scoring and statistical methodology are provided in~\Cref{app:score_computation}.

\paragraph{Implementation.}
We implement our message-passing protocol, as outlined in \Cref{sec:mp_protocol}, using LangChain  \citep{Chase_LangChain_2022} as it provides integrations with most available LLMs. 
We implement graph generation with NetworkX \footnote{\url{https://networkx.org}}, a widely used framework for graph and network processing. Our implementation is designed to be easily extensible to other graph distributions, graph sizes, and new LLMs.
We provide open-source code at \url{https://github.com/floriangroetschla/AgentsNet} and our dataset at \url{https://huggingface.co/datasets/disco-eth/AgentsNet}.

\begin{figure}
    \centering
    \includegraphics[width=1.0\linewidth]{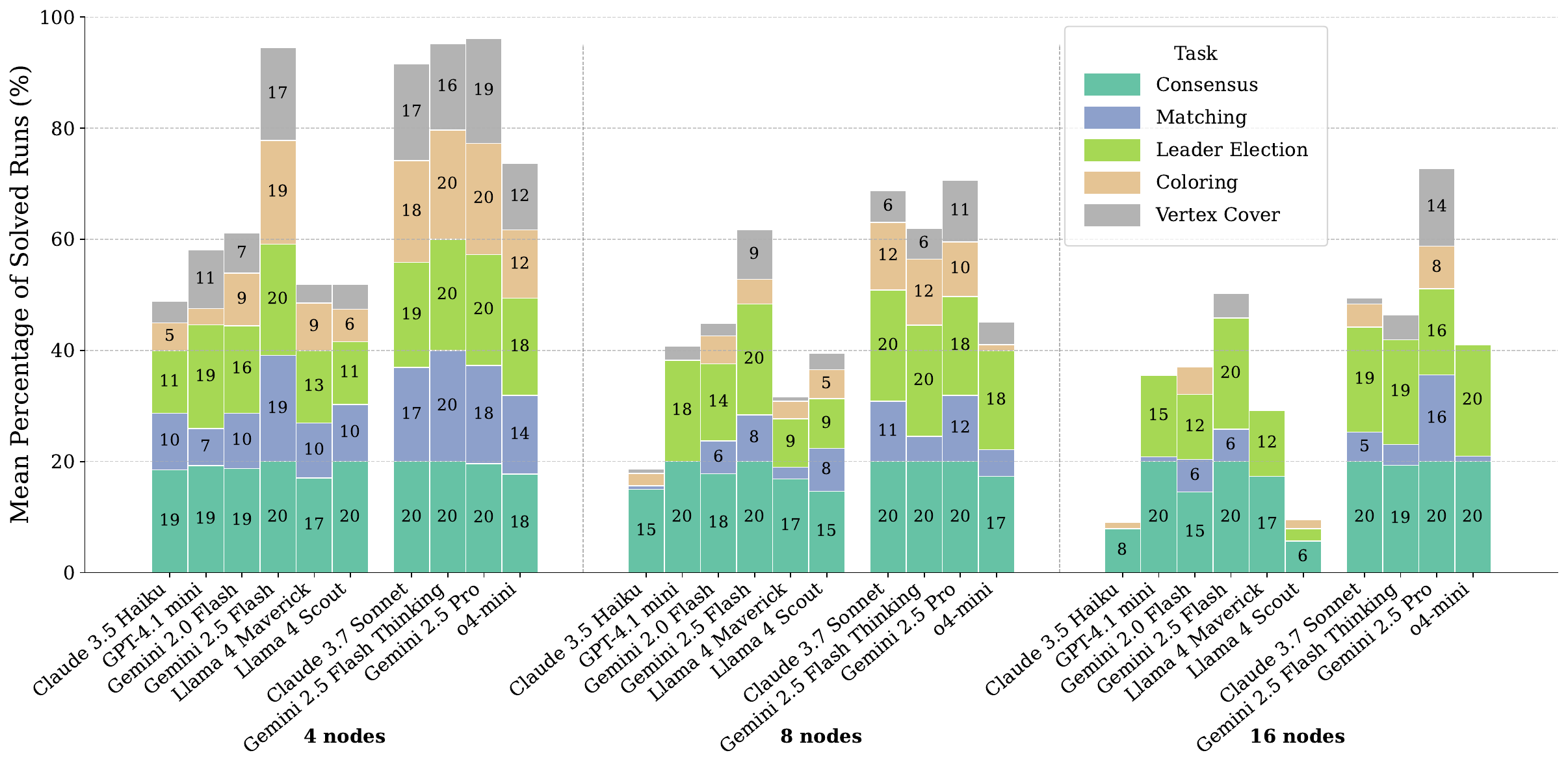}
    \caption{Fraction of solved instances per task and model, grouped by graph size (4, 8, and 16 nodes). Each task contributes up to 20\% to the total, as tasks are equally distributed across the five benchmark tasks. Reasoning and non-reasoning models are visually separated. This breakdown complements \Cref{fig:performance_by_cost} by providing a more granular view of task-level performance.}
    \label{fig:main_results}
\end{figure}

\subsection{Results on \benchmark{}}
We provide the fraction of solved instances per task in \Cref{tab:main_results}. We follow the suggestion of~\citet{miller2024adding} and report the standard error of the mean for our results. 
In addition, we plot a breakdown over different graph sizes in \Cref{fig:main_results}. Finally, in \Cref{fig:performance_by_cost} we plot the performance of models across all tasks with respect to API costs.
We observe that even for the 4-node graphs, no model performs consistently strongly across all tasks. In particular, the \textsc{Consensus} task is solved by most models, while performance on \textsc{VertexCover} is low for most models, in particular for 8 and 16 nodes. Overall, the best performing models are Claude 3.7 Sonnet, Gemini 2.5 Pro, and Gemini 2.5 Flash. In fact, Gemini 2.5 Flash is roughly on par with Claude 3.7 while being much cheaper to run on \benchmark{} (by about a factor of 20).
Interestingly, model performance generally drops with an increase in graph size. Next, we show an ablation study on further scaling the graph size to probe whether \benchmark{} can be scaled jointly with the increase in future model capabilities.

\subsection{Scaling the Agent Network}
\begin{wrapfigure}{r}{0.6\linewidth}
    \centering
    \vspace{-0.5cm}
    \includegraphics[trim={2.5cm 0 2cm 0},clip,width=\linewidth]{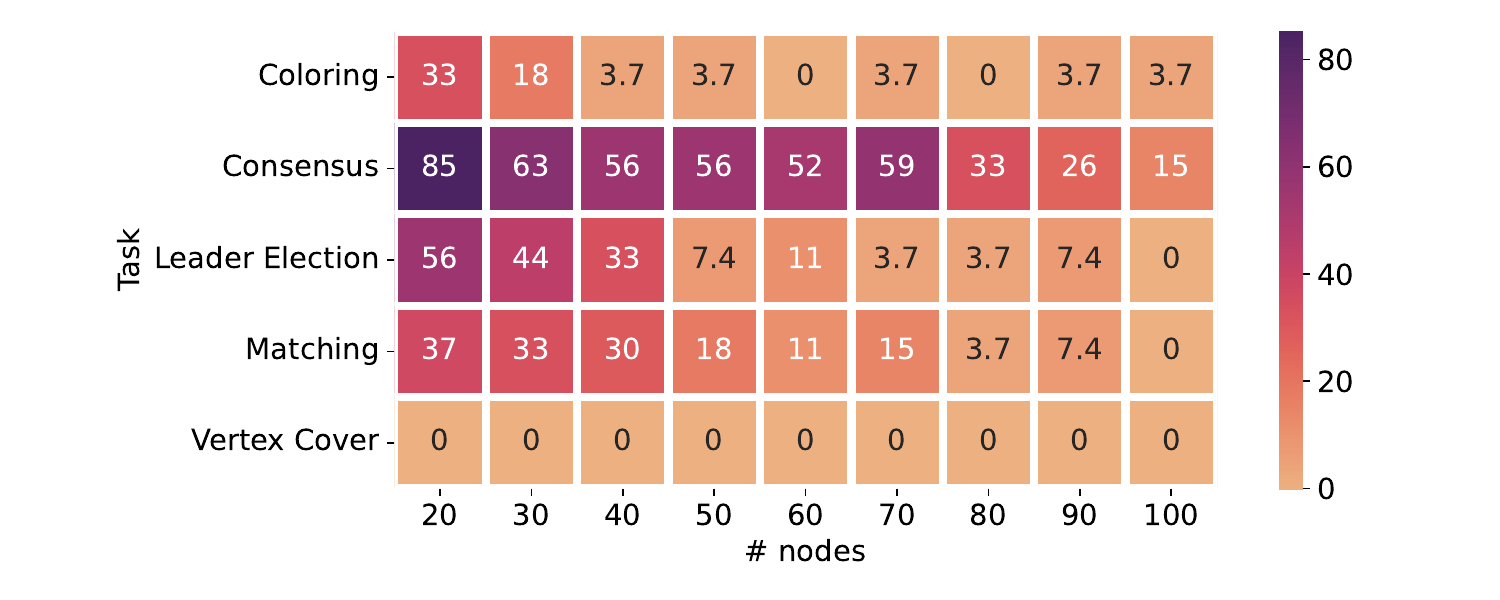}
    \caption{Scalability of Gemini 2.0 Flash on \benchmark{}: Average fraction of successfully solved instances per task as the graph size increases from 20 to 100 agents.}
    \label{fig:ablations}
\end{wrapfigure}
In addition to our main results, we provide additional results for networks of up to 100 agents in \Cref{fig:ablations} on Gemini 2.0 Flash, which shows good performance on \benchmark{} while remaining cost-efficient. Concretely, we generate a total of 81 network topologies. For simplicity, and as a good rule-of-thumb, we run message-passing for $2D + 1$ rounds, where $D$ is the graph diameter, for all tasks.
We observe that performance smoothly decreases as the network grows in size. 
Although the five tasks vary in inherent difficulty, for example, \textsc{Matching} and \textsc{Coloring} are often easier on small graphs than \textsc{Consensus} or \textsc{LeaderElection}, we observe that all tasks become substantially more challenging as the size of the network increases.
For 100-agent networks, performance drops to near zero across the board. As a consequence, the difficulty of \benchmark{} can be gradually increased by considering larger networks. Importantly, this increase in difficulty can be facilitated without any changes to \benchmark{}, which we design to allow for an arbitrary network size.

\subsection{Qualitative Analysis}
Here, we present a qualitative analysis of the responses of different LLMs to gain a deeper understanding of their overall communication, solution strategies, and collaborative capabilities. In particular, we analyze transcript data for select models across different levels of performance on \benchmark{}.  Concretely, we select Llama Maverick, Gemini 2.5 Flash, Gemini 2.5 Pro, as well as o4-mini. Here, we highlight key findings and show select examples. In \Cref{app:qualitative}, we present the full analysis and a number of examples and excerpts from transcripts. Our key findings are:

\begin{chatlog}
    \textbf{Finding 1:} Strategy coordination poses an essential challenge on \benchmark{}.
    \vspace{7px}
\end{chatlog}

We find multiple failure cases due to issues with coordinating a strategy between agents. In some cases, agents agree on a common strategy too late during message-passing, leaving an insufficient number of message-passing rounds to implement the strategy. 
In other cases, agents do not coordinate their strategy at all. Concretely, agents assume some strategy in their initial chain-of-thought and then follow that strategy throughout message-passing without informing neighbors about their strategy.%

\begin{chatlog}
    \textbf{Finding 2:} Agents generally accept information sent by neighbors.
    \vspace{7px}
\end{chatlog}

This includes key information about the network, proposed strategies, or candidate solutions. While generally enabling effective coordination, agents sometimes fail to question erroneous information, leading to incorrect solutions. Examples of such erroneous information are incorrect assumptions about the network topology or ineffective strategies proposed by other agents.

\begin{chatlog}
    \textbf{Finding 3:} Agents help their neighbors resolving inconsistencies in candidate solutions.
    \vspace{7px}
\end{chatlog}

We find multiple examples where agents detect conflicting color assignments in \textsc{Coloring} problems between other agents and assist in resolving these conflicts.
We present detailed examples and failure cases in \Cref{app:qualitative}.

\section{Limitations}\label{sec:limitations}
We implement \benchmark{} on the LOCAL computation model from distributed computing, which uses synchronous message-passing. However, other message-passing protocols may be more effective in the context of collaborative problem solving. For example, a downside of synchronous message-passing is that agents receive answers to their messages only in the subsequent message-passing round. Moreover, our implementation relies on answer parsing via JSON, which can be ambiguous and error-prone. However, we note that both of the above limitations are implementation choices not inherent to \benchmark{}. For example, future work on effective message-passing protocols or the use of structured output\footnote{\url{https://openai.com/index/introducing-structured-outputs-in-the-api/}} could be easily incorporated into \benchmark{}. 
We elaborate on the limitations in \Cref{app:limitations}.

\section{Conclusion}
In this work, we proposed \benchmark{}, a multi-agent benchmark built on top of fundamental problems from distributed computing, with the goal of assessing the ability of agentic networks to coordinate and collaborate to solve problems. 
While existing benchmarks are limited to 2--5 agents, the initial \benchmark{} suite probes up to 100 agents and is practically unlimited in size and can generate problems of increasing complexity to keep up with new generations of frontier models.
To this end, we design a robust message-passing protocol to enable multi-step communication between agents and evaluate models on a variety of graph instances, sampled from multiple graph models, and with different graph sizes. We evaluate and compare a variety of frontier LLMs on \benchmark{} and find that our tasks can be challenging for even the best models.%

\bibliographystyle{apalike}
\bibliography{bibliography}

\begin{thebibliography}{}

\bibitem[Agashe et~al., 2024]{agashe2024llm}
Agashe, S., Fan, Y., Reyna, A., and Wang, X.~E. (2024).
\newblock Llm-coordination: evaluating and analyzing multi-agent coordination
  abilities in large language models.
\newblock {\em arXiv preprint arXiv:2310.03903}.

\bibitem[Angluin, 1980]{angluin1980local}
Angluin, D. (1980).
\newblock Local and global properties in networks of processors.
\newblock In {\em Proceedings of the twelfth annual ACM symposium on Theory of
  computing}, pages 82--93.

\bibitem[Anthropic, 2024]{anthropic2024claude3}
Anthropic (2024).
\newblock The claude 3 model family: Opus, sonnet, haiku.

\bibitem[Barab{\'a}si and Albert, 1999]{barabasi1999emergence}
Barab{\'a}si, A.-L. and Albert, R. (1999).
\newblock Emergence of scaling in random networks.
\newblock {\em science}, 286(5439):509--512.

\bibitem[Barenboim, 2016]{barenboim2016deterministic}
Barenboim, L. (2016).
\newblock Deterministic ($\delta$+ 1)-coloring in sublinear (in $\delta$) time
  in static, dynamic, and faulty networks.
\newblock {\em Journal of the ACM (JACM)}, 63(5):1--22.

\bibitem[Chase, 2022]{Chase_LangChain_2022}
Chase, H. (2022).
\newblock {LangChain}.

\bibitem[Chen et~al., 2023]{chen2023multi}
Chen, H., Ji, W., Xu, L., and Zhao, S. (2023).
\newblock Multi-agent consensus seeking via large language models.
\newblock {\em arXiv preprint arXiv:2310.20151}.

\bibitem[Chen et~al., 2021]{chen2021codex}
Chen, M., Tworek, J., Jun, H., Yuan, Q., de~Oliveira~Pinto, H.~P., Kaplan, J.,
  Edwards, H., Burda, Y., Joseph, N., Brockman, G., Ray, A., Puri, R., Krueger,
  G., Petrov, M., Khlaaf, H., Sastry, G., Mishkin, P., Chan, B., Gray, S.,
  Ryder, N., Pavlov, M., Power, A., Kaiser, L., Bavarian, M., Winter, C.,
  Tillet, P., Such, F.~P., Cummings, D., Plappert, M., Chantzis, F., Barnes,
  E., Herbert-Voss, A., Guss, W.~H., Nichol, A., Paino, A., Tezak, N., Tang,
  J., Babuschkin, I., Balaji, S., Jain, S., Saunders, W., Hesse, C., Carr,
  A.~N., Leike, J., Achiam, J., Misra, V., Morikawa, E., Radford, A., Knight,
  M., Brundage, M., Murati, M., Mayer, K., Welinder, P., McGrew, B., Amodei,
  D., McCandlish, S., Sutskever, I., and Zaremba, W. (2021).
\newblock Evaluating large language models trained on code.

\bibitem[Chen et~al., 2024a]{chen2024agentverse}
Chen, W., Su, Y., Zuo, J., Yang, C., Yuan, C., Chan, C.-M., Yu, H., Lu, Y.,
  Hung, Y.-H., Qian, C., Qin, Y., Cong, X., Xie, R., Liu, Z., Sun, M., and
  Zhou, J. (2024a).
\newblock Agentverse: Facilitating multi-agent collaboration and exploring
  emergent behaviors.
\newblock In {\em The Twelfth International Conference on Learning
  Representations}.

\bibitem[Chen et~al., 2024b]{chen2024internet}
Chen, W., You, Z., Li, R., Guan, Y., Qian, C., Zhao, C., Yang, C., Xie, R.,
  Liu, Z., and Sun, M. (2024b).
\newblock Internet of agents: Weaving a web of heterogeneous agents for
  collaborative intelligence.
\newblock {\em arXiv preprint arXiv:2407.07061}.

\bibitem[Chiang et~al., 2024]{chiang2024adaptive}
Chiang, Y.-S., Cho, H.-C., and Chang, C.-J. (2024).
\newblock Adaptive networks driven by partner choice can facilitate
  coordination among humans in the graph coloring game: Evidence from a network
  experiment.
\newblock {\em Collective Intelligence}, 3(3):26339137241285901.

\bibitem[Chuang et~al., 2024]{chuang2024simulating}
Chuang, Y.-S., Goyal, A., Harlalka, N., Suresh, S., Hawkins, R., Yang, S.,
  Shah, D., Hu, J., and Rogers, T. (2024).
\newblock Simulating opinion dynamics with networks of {LLM}-based agents.
\newblock In {\em Findings of the Association for Computational Linguistics:
  NAACL 2024}.

\bibitem[Du et~al., 2023]{du2023improving}
Du, Y., Li, S., Torralba, A., Tenenbaum, J.~B., and Mordatch, I. (2023).
\newblock Improving factuality and reasoning in language models through
  multiagent debate.
\newblock In {\em Forty-first International Conference on Machine Learning}.

\bibitem[Erdos et~al., 1960]{erdos1960evolution}
Erdos, P., R{\'e}nyi, A., et~al. (1960).
\newblock On the evolution of random graphs.
\newblock {\em Publ. math. inst. hung. acad. sci}, 5(1):17--60.

\bibitem[Fatemi et~al., 2024]{fatemi2024talk}
Fatemi, B., Halcrow, J., and Perozzi, B. (2024).
\newblock Talk like a graph: Encoding graphs for large language models.
\newblock In {\em International Conference on Learning Representations (ICLR)}.

\bibitem[Fischer et~al., 1985]{fischer1985impossibility}
Fischer, M.~J., Lynch, N.~A., and Paterson, M.~S. (1985).
\newblock Impossibility of distributed consensus with one faulty process.
\newblock {\em Journal of the ACM (JACM)}, 32(2):374--382.

\bibitem[Google, 2024]{google2024gemini2.0}
Google (2024).
\newblock Gemini 2.0: A new ai model for the agentic era.

\bibitem[Google, 2025a]{google2024gemini2.5flash}
Google (2025a).
\newblock Developers can now start building with gemini 2.5 flash.

\bibitem[Google, 2025b]{google2024gemini2.5pro}
Google (2025b).
\newblock Gemini 2.5: Our newest gemini model with thinking.

\bibitem[Hendrycks et~al., 2021]{hendrycks_mmlu}
Hendrycks, D., Burns, C., Basart, S., Zou, A., Mazeika, M., Song, D., and
  Steinhardt, J. (2021).
\newblock Measuring massive multitask language understanding.
\newblock {\em Proceedings of the International Conference on Learning
  Representations (ICLR)}.

\bibitem[Judd et~al., 2010]{judd2010behavioral}
Judd, S., Kearns, M., and Vorobeychik, Y. (2010).
\newblock Behavioral dynamics and influence in networked coloring and
  consensus.
\newblock {\em Proceedings of the National Academy of Sciences},
  107(34):14978--14982.

\bibitem[Kearns et~al., 2006]{kearns2006experimental}
Kearns, M., Suri, S., and Montfort, N. (2006).
\newblock An experimental study of the coloring problem on human subject
  networks.
\newblock {\em science}, 313(5788):824--827.

\bibitem[Lenzen and Wattenhofer, 2012]{lenzen2012distributed}
Lenzen, C. and Wattenhofer, R. (2012).
\newblock Distributed algorithms for sensor networks.
\newblock {\em Philosophical Transactions of the Royal Society A: Mathematical,
  Physical and Engineering Sciences}, 370(1958):11--26.

\bibitem[Liang et~al., 2024]{liang2024encouraging}
Liang, T., He, Z., Jiao, W., Wang, X., Wang, Y., Wang, R., Yang, Y., Shi, S.,
  and Tu, Z. (2024).
\newblock Encouraging divergent thinking in large language models through
  multi-agent debate.
\newblock In {\em Proceedings of the 2024 Conference on Empirical Methods in
  Natural Language Processing}, pages 17889--17904.

\bibitem[Linial, 1992]{linial1992local}
Linial, N. (1992).
\newblock Locality in distributed graph algorithms.
\newblock {\em SIAM Journal on computing}, 21(1):193--201.

\bibitem[Liu et~al., 2024]{liu2024agentbench}
Liu, X., Yu, H., Zhang, H., Xu, Y., Lei, X., Lai, H., Gu, Y., Ding, H., Men,
  K., Yang, K., Zhang, S., Deng, X., Zeng, A., Du, Z., Zhang, C., Shen, S.,
  Zhang, T., Su, Y., Sun, H., Huang, M., Dong, Y., and Tang, J. (2024).
\newblock Agentbench: Evaluating {LLM}s as agents.
\newblock In {\em The Twelfth International Conference on Learning
  Representations}.

\bibitem[Liu et~al., 2023]{liu2023dynamic}
Liu, Z., Zhang, Y., Li, P., Liu, Y., and Yang, D. (2023).
\newblock Dynamic llm-agent network: An llm-agent collaboration framework with
  agent team optimization.
\newblock {\em arXiv preprint arXiv:2310.02170}.

\bibitem[Lynch, 1996]{lynch1996distributed}
Lynch, N.~A. (1996).
\newblock {\em Distributed algorithms}.
\newblock Elsevier.

\bibitem[Marro et~al., 2024]{marro2024scalable}
Marro, S., La~Malfa, E., Wright, J., Li, G., Shadbolt, N., Wooldridge, M., and
  Torr, P. (2024).
\newblock A scalable communication protocol for networks of large language
  models.
\newblock {\em arXiv preprint arXiv:2410.11905}.

\bibitem[Meta, 2025]{meta2025llama4}
Meta (2025).
\newblock The llama 4 herd: The beginning of a new era of natively multimodal
  ai innovation.

\bibitem[Mialon et~al., 2024]{mialon2024gaia}
Mialon, G., Fourrier, C., Wolf, T., LeCun, Y., and Scialom, T. (2024).
\newblock {GAIA}: a benchmark for general {AI} assistants.
\newblock In {\em The Twelfth International Conference on Learning
  Representations}.

\bibitem[Miller, 2024]{miller2024adding}
Miller, E. (2024).
\newblock Adding error bars to evals: A statistical approach to language model
  evaluations.
\newblock {\em arXiv preprint arXiv:2411.00640}.

\bibitem[Ni et~al., 2025]{ni2025coral}
Ni, A., Desai, R., Li, Y., Lei, X., Wang, D., Raghavendra, R., Ghosh, G., Li,
  D., and Celikyilmaz, A. (2025).
\newblock Collaborative reasoner: Self-improving social agents with synthetic
  conversations.
\newblock {\em arXiv preprint}.

\bibitem[OpenAI, 2025a]{openai2025gpt4.1}
OpenAI (2025a).
\newblock Introducing gpt-4.1 in the api.

\bibitem[OpenAI, 2025b]{openai2025o4}
OpenAI (2025b).
\newblock Introducing openai o3 and o4-mini.

\bibitem[Park et~al., 2023]{park2023generative}
Park, J.~S., O'Brien, J., Cai, C.~J., Morris, M.~R., Liang, P., and Bernstein,
  M.~S. (2023).
\newblock Generative agents: Interactive simulacra of human behavior.
\newblock In {\em Proceedings of the 36th annual acm symposium on user
  interface software and technology}, pages 1--22.

\bibitem[Peleg, 2000]{peleg2000distributed}
Peleg, D. (2000).
\newblock {\em Distributed computing: a locality-sensitive approach}.
\newblock SIAM.

\bibitem[Qian et~al., 2024]{qian2024scaling}
Qian, C., Xie, Z., Wang, Y., Liu, W., Dang, Y., Du, Z., Chen, W., Yang, C.,
  Liu, Z., and Sun, M. (2024).
\newblock Scaling large-language-model-based multi-agent collaboration.
\newblock {\em arXiv preprint arXiv:2406.07155}.

\bibitem[Regan et~al., 2024]{regan2024problem}
Regan, C., Gournail, A., and Oka, M. (2024).
\newblock Problem-solving in language model networks.
\newblock In {\em Artificial Life Conference Proceedings 36}.

\bibitem[Sanford et~al., 2024]{sanford2024understanding}
Sanford, C., Fatemi, B., Hall, E., Tsitsulin, A., Kazemi, M., Halcrow, J.,
  Perozzi, B., and Mirrokni, V. (2024).
\newblock Understanding transformer reasoning capabilities via graph
  algorithms.
\newblock In {\em The Thirty-eighth Annual Conference on Neural Information
  Processing Systems}.

\bibitem[Schaeffer et~al., 2023]{schaeffer2023emergent}
Schaeffer, R., Miranda, B., and Koyejo, S. (2023).
\newblock Are emergent abilities of large language models a mirage?
\newblock In {\em NeurIPS}.

\bibitem[Skianis et~al., 2024]{skianis2024graphreasoninglargelanguage}
Skianis, K., Nikolentzos, G., and Vazirgiannis, M. (2024).
\newblock Graph reasoning with large language models via pseudo-code prompting.

\bibitem[Tang et~al., 2025]{tang2025evaluating}
Tang, J., Zhang, Q., Li, Y., Chen, N., and Li, J. (2025).
\newblock Evaluating and improving large language models on graph computation.
\newblock In {\em The Thirteenth International Conference on Learning
  Representations}.

\bibitem[Wang et~al., 2024]{wang2024can}
Wang, H., Feng, S., He, T., Tan, Z., Han, X., and Tsvetkov, Y. (2024).
\newblock Can language models solve graph problems in natural language?
\newblock {\em Advances in Neural Information Processing Systems}, 36.

\bibitem[Watts and Strogatz, 1998]{watts1998collective}
Watts, D.~J. and Strogatz, S.~H. (1998).
\newblock Collective dynamics of ‘small-world’networks.
\newblock {\em nature}, 393(6684):440--442.

\bibitem[Xiong et~al., 2023]{xiong2023examining}
Xiong, K., Ding, X., Cao, Y., Liu, T., and Qin, B. (2023).
\newblock Examining inter-consistency of large language models collaboration:
  An in-depth analysis via debate.
\newblock In {\em The 2023 Conference on Empirical Methods in Natural Language
  Processing}.

\bibitem[Xu et~al., 2023]{xu2023magic}
Xu, L., Hu, Z., Zhou, D., Ren, H., Dong, Z., Keutzer, K., Ng, S.~K., and Feng,
  J. (2023).
\newblock Magic: Benchmarking large language model powered multi-agent in
  cognition, adaptability, rationality and collaboration.
\newblock {\em arXiv preprint arXiv:2311.08562}.

\bibitem[Yang et~al., 2024]{yang2024oasisopenagentsocial}
Yang, Z., Zhang, Z., Zheng, Z., Jiang, Y., Gan, Z., Wang, Z., Ling, Z., Chen,
  J., Ma, M., Dong, B., Gupta, P., Hu, S., Yin, Z., Li, G., Jia, X., Wang, L.,
  Ghanem, B., Lu, H., Lu, C., Ouyang, W., Qiao, Y., Torr, P., and Shao, J.
  (2024).
\newblock Oasis: Open agent social interaction simulations with one million
  agents.

\bibitem[Yao et~al., 2024]{yao2024tau}
Yao, S., Shinn, N., Razavi, P., and Narasimhan, K. (2024).
\newblock Tau-bench: A benchmark for tool-agent-user interaction in real-world
  domains.
\newblock {\em arXiv preprint arXiv:2406.12045}.

\bibitem[Yin et~al., 2024]{yin2024mmau}
Yin, G., Bai, H., Ma, S., Nan, F., Sun, Y., Xu, Z., Ma, S., Lu, J., Kong, X.,
  Zhang, A., et~al. (2024).
\newblock Mmau: A holistic benchmark of agent capabilities across diverse
  domains.
\newblock {\em arXiv preprint arXiv:2407.18961}.

\bibitem[Zhang et~al., 2024]{zhang2024llm4dyg}
Zhang, Z., Wang, X., Zhang, Z., Li, H., Qin, Y., and Zhu, W. (2024).
\newblock Llm4dyg: Can large language models solve spatial-temporal problems on
  dynamic graphs?
\newblock In {\em Proceedings of the 30th ACM SIGKDD Conference on Knowledge
  Discovery and Data Mining}, KDD '24, page 4350–4361, New York, NY, USA.
  Association for Computing Machinery.

\bibitem[Zhuge et~al., 2023]{zhuge2023mindstorms}
Zhuge, M., Liu, H., Faccio, F., Ashley, D.~R., Csord{\'{a}}s, R.,
  Gopalakrishnan, A., Hamdi, A., Hammoud, H. A. A.~K., Herrmann, V., Irie, K.,
  Kirsch, L., Li, B., Li, G., Liu, S., Mai, J., Piekos, P., Ramesh, A.~A.,
  Schlag, I., Shi, W., Stanic, A., Wang, W., Wang, Y., Xu, M., Fan, D., Ghanem,
  B., and Schmidhuber, J. (2023).
\newblock Mindstorms in natural language-based societies of mind.
\newblock {\em Arxiv}.

\bibitem[Zhuge et~al., 2024]{zhuge2024gptswarm}
Zhuge, M., Wang, W., Kirsch, L., Faccio, F., Khizbullin, D., and Schmidhuber,
  J. (2024).
\newblock Gptswarm: Language agents as optimizable graphs.
\newblock In {\em ICML}.

\end{thebibliography}

\newpage

\newpage
\appendix

\section{Implementation Details}\label{app:implementation}
Here, we describe implementation details of \benchmark{}.

\paragraph{Message-Passing.}
\Cref{algo:mp_protocol} gives an overview over our message-passing in pseudocode. We generate a message from agent $v$ with $\textsc{\texttt{Generate}}(v \mid P)$, where $P$ is a information provided in the prompt. Agent $v$ can send/receive messages to/from neighbors $w \in N(v)$ with $\textsc{\texttt{SendMessage}}(m, w)$ and $\textsc{\texttt{ReceiveMessage}}(w)$, respectively. For clarity, we omit the re-tries and JSON parsing from \Cref{algo:mp_protocol}.

\begin{algorithm}
\caption{Pseudocode for $T$ rounds of message-passing.}\label{algo:mp_protocol}
\begin{algorithmic}
\For{\textbf{each} agent $v$}
\vspace{2px}
    \State $m \gets \textsc{\texttt{Generate}}(v \mid \textbf{System prompt})$ %
\vspace{2px}
    \State \textbf{for each} neighbor $w$ \textbf{do}: $\textsc{\texttt{SendMessage}}(m, w)$
\EndFor
\vspace{5px}
\For{\textbf{each }$t \in \{1, \dots, T-1\}$}
\vspace{2px}
    \State \textbf{for each} neighbor $w$ \textbf{do}: $m(w) \gets \textsc{\texttt{ReceiveMessage}}(w)$
\vspace{2px}
\State $m \gets \textsc{\texttt{Generate}}(v \mid \textbf{for each } \text{neighbor } $w$: m(w))$
\vspace{2px}
\State \textbf{for each} neighbor $w$ \textbf{do}: $\textsc{\texttt{SendMessage}}(m, w)$
\EndFor
\vspace{5px}
\State \Return \textbf{for each} agent $v$: $\textsc{\texttt{Generate}}(v \mid \textbf{Result prompt})$
\end{algorithmic}
\end{algorithm}

\begin{table}[b]
    \centering
    \resizebox{\textwidth}{!}{%
    \begin{tabular}{lll}
    \toprule
    Model & Provider & Version \\
    \midrule
    Claude 3.5 Haiku  & Anthropic & \texttt{claude-3-5-haiku-20241022} \\
    Claude 3.7 Sonnet  & Anthropic & \texttt{claude-3-7-sonnet-20250219} \\
    GPT-4.1 mini  & OpenAI & \texttt{gpt-4.1-mini} \\
    o4-mini  & OpenAI & \texttt{o4-mini} \\
    Gemini 2.0 Flash  & Google & \texttt{gemini-2.0-flash} \\
    Gemini 2.5 Flash  & Google & \texttt{gemini-2.5-flash-preview-04-17} \\
    Gemini 2.5 FT  & Google & \texttt{gemini-2.5-flash-preview-04-17-thinking} \\
    \multirow{ 2}{*}{Gemini 2.5 Pro}  & \multirow{ 2}{*}{Google} & \texttt{gemini-2.5-pro-preview-03-25} and  \\
    & & \texttt{gemini-2.5-pro-preview-05-06} \\
    Llama 4 Maverick  & Together AI & \texttt{meta-llama/Llama-4-Maverick-17B-128E-Instruct-FP8} \\
    Llama 4 Scout  & Together AI & \texttt{meta-llama/Llama-4-Scout-17B-16E-Instruct} \\
    \bottomrule
    \end{tabular}%
    }
    \caption{API providers and model versions used in our testing.}
    \label{tab:api}
\end{table}

\paragraph{Models.}
We provide details on API providers and model versions in \Cref{tab:api}, which includes a diverse range of proprietary and open-source LLMs that span instruction-tuned, reasoning-enhanced, and cost-efficient models. These were selected to ensure a broad coverage of state-of-the-art capabilities, as well as compatibility with the long-context requirements of \benchmark{} tasks. All models included support effective context lengths exceeding 16k tokens, which is essential for multi-round communication in graphs with up to 16 nodes.

\paragraph{Full System Prompt.}
Here, we provide the full system prompt for \benchmark{}. Since the prompt differs between tasks, agent, and the number of agents in the network, we use template variables, which we indicate here with \texttt{[n]} (number of agents), \texttt{[name]} (name of agent), \texttt{[neighbors]} (comma-separated names of neighbors), \texttt{[r]} (number of message-passing rounds), \texttt{[task1]} (first part of the task description, at the beginning of the prompt), and \texttt{[task2]} (second part of the task description, at the end of the prompt). We provide the task-specific descriptions \texttt{[task1]} and \texttt{[task2]} for each task in \Cref{app:benchmark_tasks}.

\setChatlogWidth{\textwidth}
\begin{chatlog}
\chatmsg[1]{System}{\texttt{\small You are an agent that is connected with other agents (your neighbors), who you communicate with. Your neighbors can in turn communicate with their neighbors and so forth. \texttt{[task1]}.
The rules are as follows:
\begin{enumerate}
    \item There are \texttt{[n]} agents in total. Everybody has a unique name. Your name is \texttt{[name]}.
    \item You can only communicate with your immediate neighbors (\texttt{[neighbors]}). You cannot see or directly communicate with anyone else, unless information is relayed by intermediate agents.
    \item You can exchange text-based messages with your neighbors in rounds. In each round, you will first receive the last messages sent by your neighbors and then be asked to generate your response messages which your neighbors receive in the next round. This process repeats for \texttt{[r]} rounds of message passing. Importantly, the process is synchronous: Every agent decides on which messages to send at the same time and sees the messages from other agents only in the next round.
    \item Everybody (including you) decides what to share or request from neighbors. In every round, think step-by-step about the next set of messages you want to send. Output a JSON string that contains your response messages.
    \item The messages you send to your neighbors are formatted as JSON. For example, if your neighbors are Alan and Bob, your output should look as follows: \texttt{`}\texttt{`}\texttt{`} \{"Alan": "Message that will be sent to Alan.", "Bob": "Message that will be sent to Bob."\} \texttt{`}\texttt{`}\texttt{`} It is not mandatory to send a message to every neighbor in every round. If you do not want to send a message to a particular neighbor, you may omit their name from the JSON.
    \item After \texttt{[r]} message passes, you have to solve the following task: \texttt{[task2]}.
\end{enumerate}
}}
\end{chatlog}

\section{Benchmark Tasks}\label{app:benchmark_tasks}
Here, we describe the tasks in \benchmark{} in detail.

\paragraph{($\Delta + 1$)-Coloring.} Each node is assigned a color using at most $\Delta + 1$ colors, where $\Delta$ is the maximum node degree. This problem has a well-defined distributed complexity of $O(\log^* n)$ in bounded-degree graphs~\citep{barenboim2016deterministic}.
This task is particularly useful for role assignment within multi-agent systems. For instance, agents can be designated to perform specific sub-tasks (e.g., web search, reasoning, coding, planning), with the constraint that directly connected agents are assigned distinct roles to avoid redundancy. Solving this task reflects the system's ability to efficiently distribute responsibilities across the network with minimal overlap in capabilities.

The corresponding agentic task is to form groups, with a pre-defined number of groups and where each group corresponds to a color. After message-passing, each agent is asked to respond with the group it wants to be in. The evaluation score is designed to reflect the number of connected agents in the same group.
Let $A(u)$ denote answer of agent $u$, then the score is computed as
\begin{equation*}
    \dfrac{\sum_{(u, v) \in \textbf{edges}} \mathbf{1}(A(u) \neq A(v))}{\#\textbf{edges}},
\end{equation*}
where $\mathbf{1}(x) = 1$ if $x$ is true and $0$ otherwise. In \benchmark{}, we refer to this task as \textsc{Coloring} and provide the following task descriptions.

\setChatlogWidth{\textwidth}
\begin{chatlog}
\chatmsg[1]{\texttt{[task1]}}{\texttt{\small Your task is to partition yourselves into groups such that agents who are neighbors are never in the same group.}
}
\end{chatlog}

\setChatlogWidth{\textwidth}
\begin{chatlog}
\chatmsg[1]{\texttt{[task2]}}{\texttt{\small You will be requested to state which group you assign yourself to. There are exactly \texttt{[$\Delta + 1$]} groups available: Group 1,...,Group \texttt{[$\Delta + 1$]}. You should assign yourself to exactly one of these groups. The final result should be such that any two agents who are neighbors are in different groups. In particular, you should assign yourself to a group that is different from all of your neighbors' groups. }
}
\end{chatlog}

Note that \texttt{[$\Delta + 1$]} is a template variable resolving to one plus the maximum degree of the network.

\paragraph{Minimal Vertex Cover.} A minimal vertex cover is a subset of nodes such that every edge in the graph has at least one endpoint in the subset, and removing any node from this subset would violate that property. This problem has a close relationship with the maximal independent set and is similarly fundamental in distributed computing, with known randomized solutions in $O(\log n)$ rounds.
In agentic networks, a minimal vertex cover can represent a minimal set of monitor or gateway agents that maintain awareness of all interactions in the system. These agents could take on responsibilities such as relaying messages, auditing behavior, or bridging subgroups. The task tests a system's ability to identify a compact yet effective set of nodes with high influence or observability.

The corresponding agentic task is to select a group of coordinators among the agents. After message-passing, each agent is asked to indicate whether it is a coordinator. The agents can respond with either \textit{Yes} or \textit{No}. The evaluation score is designed to reflect both the ratio of connected agents at least one of which is a coordinator, as well as the number of times the minimality constraint is violated. Let $A(u)$ denote the answer of agent $u$, we first compute the ratio of covered edges as
\begin{equation*}
    \textbf{coverage} := \dfrac{\sum_{(u, v) \in \textbf{edges}} \mathbf{1}(A(u) = \textit{Yes} \vee A(v) = \textit{Yes})}{\#\textbf{edges}}.
\end{equation*}
For the minimality constraint, we count the number of \textit{non-essential} coordinators, that is, those coordinators $u$ whose neighbors are also coordinators. Each such $u$ violates the minimality constraint, as the set of coordinators without $u$ is still a vertex cover. Let $N$ denote the number of non-essential coordinators, then the evaluation score is computed as
\begin{equation*}
    \textbf{coverage} \cdot \Big(1 - \frac{N}{\#\textbf{coordinators}}\Big).
\end{equation*}
In \benchmark{}, we refer to this task as \textsc{VertexCover} and provide the following task descriptions.

\setChatlogWidth{\textwidth}
\begin{chatlog}
\chatmsg[1]{\texttt{[task1]}}{\texttt{\small Your task is to select, among all agents, a group of coordinators such that whenever two agents communicate at least one of them is a coordinator. The group of coordinators should be selected such that every coordinator has at least one neighbor who is not a coordinator.}
}
\end{chatlog}

\setChatlogWidth{\textwidth}
\begin{chatlog}
\chatmsg[1]{\texttt{[task2]}}{\texttt{\small  You will be requested to state whether you are a coordinator. The response should either be 'Yes' or 'No'.}
}
\end{chatlog}

\paragraph{Maximal Matching.} A maximal matching is a set of edges such that no two edges share a vertex, and no additional edges can be added without violating this property. This task captures the ability of agents to negotiate pairwise agreements without global knowledge, which is useful in scenarios where resource allocation or mutual exclusivity must be enforced (e.g., agent-to-agent task assignment).
Randomized algorithms typically solve this problem in $O(\log n)$ rounds~\cite{peleg2000distributed}. 

The corresponding agentic task is for the agents to form pairs. After message-passing, each agent is asked to name the neighbor it wants to pair up with. The agents can also respond with \textit{None}, if they cannot find a match (all neighbor agents are already paired up with other agents). The evaluation score is designed to reflect the number of inconsistencies between agents. Possible inconsistencies are: (a) agent $u$ selected agent $v$ but agent $v$ did not select agent $u$; (b) Agent $u$ selected an agent that $u$ is not connected to; (c) agent $u$ answered \textit{None}, but there is an agent $v$ that is a neighbor of $u$ which also answered \textit{None}, meaning that $u$ and $v$ could form a pair. Let $I$ denote the number of inconsistencies, then the evaluation score is computed as
\begin{equation*}
    1 - \frac{I}{\#\mathbf{agents}}.
\end{equation*}
In \benchmark{}, we refer to this task as \textsc{Matching} and provide the following task descriptions.

\setChatlogWidth{\textwidth}
\begin{chatlog}
\chatmsg[1]{\texttt{[task1]}}{\texttt{\small Your task is to find build groups of two agents each which can communicate with each other.}
}
\end{chatlog}

\setChatlogWidth{\textwidth}
\begin{chatlog}
\chatmsg[1]{\texttt{[task2]}}{\texttt{\small  You will be requested to name one of your neighbors that you build a group with or 'None' if all your neighbors are already assigned to other groups and cannot be in a group with you. In the end, every agent should only be in at most one group and agents in the same group have to name each other as the second group member consistently.}
}
\end{chatlog}

\paragraph{Leader Election.} One node must be selected as the leader, while all others acknowledge that they are not. This classic coordination task is central to evaluating how well agents establish hierarchy and delegate global decision-making~\citep{angluin1980local}.
In multi-agent systems, leader election can be interpreted as selecting a central planner or controller agent responsible for strategy synthesis while the remaining agents act as executors. Effective leader election demonstrates the system's capacity to break symmetry and converge on a single authority. In general graphs, the round complexity is $O(D)$, where $D$ is the network diameter.

The corresponding agentic task is to select a single leader among the agents. After message-passing, each agent is asked whether it is the leader. The agents can respond with either \textit{Yes} or \textit{No}. Let $A(u)$ denote the answer of agent $u$, then the evaluation score is computed as
\begin{equation*}
    \mathbf{1}\Big( 1 = \sum_{u \in \mathbf{agents}} \mathbf{1}(A(u) = \textit{Yes})\Big).
\end{equation*}
In \benchmark{}, we refer to this task as \textsc{LeaderElection} and provide the following task descriptions.

\setChatlogWidth{\textwidth}
\begin{chatlog}
\chatmsg[1]{\texttt{[task1]}}{\texttt{\small Your task is to collaboratively solve the problem of electing a single leader.}
}
\end{chatlog}

\setChatlogWidth{\textwidth}
\begin{chatlog}
\chatmsg[1]{\texttt{[task2]}}{\texttt{\small You will be requested to state whether or not you are the leader. The response should either be 'Yes' or 'No'. The final result should be such that exactly one agent responds with 'Yes' and all others say 'No' as there should be exactly one leader.}
}
\end{chatlog}

\paragraph{Consensus.}
In the consensus problem, all agents must agree on a single value from the set ${0, 1}$. In our benchmark, we focus on the basic setting without any faulty or Byzantine agents. The goal is for all agents to coordinate and produce the same final answer after a number of communication rounds. A successful solution requires that every agent outputs the same value, either 0 or 1. This task tests the ability of multi-agent systems to converge to a global agreement through local message-passing alone. In synchronous networks, achieving consensus generally requires $\mathcal{O}(D)$ rounds, where $D$ is the network diameter.

The corresponding agentic task is to choose between a value $0$ and $1$. After message-passing, each agent is asked to announce its selected value. Let $A(u)$ denote the answer of agent $u$, then the evaluation score is computed as
\begin{equation*}
    \mathbf{1}(\textbf{count} = \#\textbf{agents} \vee \textbf{count} = 0),
\end{equation*}
where
\begin{equation*}
    \textbf{count} := \sum_{u \in \mathbf{agents}} A(u).
\end{equation*}
In \benchmark{}, we refer to this task as \textsc{Consensus} and provide the following task descriptions.

\setChatlogWidth{\textwidth}
\begin{chatlog}
\chatmsg[1]{\texttt{[task1]}}{\texttt{\small Your goal is to agree on a single value with the other agents. The possible values that you can decide on are either 0 or 1.}
}
\end{chatlog}

\setChatlogWidth{\textwidth}
\begin{chatlog}
\chatmsg[1]{\texttt{[task2]}}{\texttt{\small After the last round, each agent must decide on a single value.}
}
\end{chatlog}

\paragraph{Results for Soft Scores}
\begin{table}[]
    \centering
    \resizebox{\textwidth}{!}{
    \begin{tabular}{lccccc}
    \toprule
       Model  & \scriptsize \textsc{Coloring} & \scriptsize \textsc{Consensus} & \scriptsize \textsc{Leader Election} & \scriptsize \textsc{Matching} & \scriptsize \textsc{Vertex Cover}  \\
       \midrule
Claude 3.5 Haiku & 0.80 \tiny \textcolor{gray}{(0.02)} & 0.69 \tiny \textcolor{gray}{(0.05)} & 0.19 \tiny \textcolor{gray}{(0.03)} & 0.69 \tiny \textcolor{gray}{(0.02)} & 0.67 \tiny \textcolor{gray}{(0.03)}\\
Claude 3.7 Sonnet & 0.96 \tiny \textcolor{gray}{(0.01)} & 1.00 \tiny \textcolor{gray}{(0.00)} & 0.96 \tiny \textcolor{gray}{(0.03)} & 0.84 \tiny \textcolor{gray}{(0.03)} & 0.85 \tiny \textcolor{gray}{(0.02)}\\
GPT-4.1 mini & 0.58 \tiny \textcolor{gray}{(0.03)} & 0.99 \tiny \textcolor{gray}{(0.01)} & 0.86 \tiny \textcolor{gray}{(0.05)} & 0.58 \tiny \textcolor{gray}{(0.03)} & 0.78 \tiny \textcolor{gray}{(0.03)}\\
Gemini 2.0 Flash & 0.86 \tiny \textcolor{gray}{(0.02)} & 0.85 \tiny \textcolor{gray}{(0.04)} & 0.69 \tiny \textcolor{gray}{(0.05)} & 0.80 \tiny \textcolor{gray}{(0.03)} & 0.75 \tiny \textcolor{gray}{(0.02)}\\
Gemini 2.5 Flash & 0.85 \tiny \textcolor{gray}{(0.03)} & 1.00 \tiny \textcolor{gray}{(0.00)} & 1.00 \tiny \textcolor{gray}{(0.00)} & 0.87 \tiny \textcolor{gray}{(0.02)} & 0.88 \tiny \textcolor{gray}{(0.03)}\\
Gemini 2.5 FT & 0.88 \tiny \textcolor{gray}{(0.03)} & 0.99 \tiny \textcolor{gray}{(0.01)} & 0.98 \tiny \textcolor{gray}{(0.02)} & 0.84 \tiny \textcolor{gray}{(0.01)} & 0.88 \tiny \textcolor{gray}{(0.02)}\\
Gemini 2.5 Pro & 0.96 \tiny \textcolor{gray}{(0.01)} & 0.99 \tiny \textcolor{gray}{(0.01)} & 0.89 \tiny \textcolor{gray}{(0.06)} & 0.93 \tiny \textcolor{gray}{(0.01)} & 0.92 \tiny \textcolor{gray}{(0.03)} \\
Llama 4 Maverick & 0.82 \tiny \textcolor{gray}{(0.02)} & 0.85 \tiny \textcolor{gray}{(0.04)} & 0.56 \tiny \textcolor{gray}{(0.06)} & 0.77 \tiny \textcolor{gray}{(0.02)} & 0.63 \tiny \textcolor{gray}{(0.03)} \\
Llama 4 Scout & 0.79 \tiny \textcolor{gray}{(0.04)} & 0.67 \tiny \textcolor{gray}{(0.05)} & 0.38 \tiny \textcolor{gray}{(0.06)} & 0.77 \tiny \textcolor{gray}{(0.02)} & 0.79 \tiny \textcolor{gray}{(0.02)}\\
o4-mini & 0.71 \tiny \textcolor{gray}{(0.03)} & 0.92 \tiny \textcolor{gray}{(0.04)} & 0.92 \tiny \textcolor{gray}{(0.03)} & 0.72 \tiny \textcolor{gray}{(0.02)} & 0.73 \tiny \textcolor{gray}{(0.02)}\\
       \bottomrule
    \end{tabular}}
    \caption{Soft scores for all tasks and models. We observe similar trends as for the fraction of solved instances. As the scores are task specific, we do not aggregate them to a total score.}
    \label{tab:app_score_results}
\end{table}
The previously defined scores are presented in \Cref{tab:app_score_results} and their computation otherwise follows the methodology used for the main scores (see \Cref{app:score_computation}). Although they generally agree with the fraction of solved instances, the numbers are harder to interpret, as even a naive baseline that predicts random values out of a set of valid responses (e.g., random colors from the valid classes for the task \textsc{Coloring} achieves scores well above 50\%. This reduces the actual range of meaningful scores to small intervals. As scores can also not be compared between tasks, we report the mean fraction of solved instances as the final benchmarking score for \benchmark{}.

\section{Score Computation}\label{app:score_computation}
We apply a systematic statistical methodology to evaluate model performance and quantify uncertainty. For each combination of graph size, task, and graph generator (Watts-Strogatz, Barabási-Albert, and Delaunay triangulations), we generate three distinct graph instances. We conduct one or more experimental runs per instance, resulting in at least three observations per configuration.
For each model, we compute a mean score $\mu_{s,t,g}$ for each configuration triplet $(s, t, g)$ where $s$ represents graph size, $t$ represents task type, and $g$ represents the graph generation algorithm:
\begin{equation}
\mu_{s,t,g} = \frac{1}{N_{s,t,g}} \sum_{i=1}^{3}\sum_{j=1}^{n_{i}} x_{s,t,g,i,j}
\end{equation}
where $x_{s,t,g,i,j}$ denotes the performance score of the $j$-th run on the $i$-th graph instance of configuration $(s, t, g)$, $n_{i}$ is the number of runs performed on the $i$-th graph instance, and $N_{s,t,g} = \sum_{i=1}^{3} n_{i}$ is the total number of runs for this configuration.
For each configuration, we compute the standard error $\text{SE}_{s,t,g}$ as:
\begin{equation}
\text{SE}_{s,t,g} = \frac{\sigma_{s,t,g}}{\sqrt{N_{s,t,g}}}
\end{equation}
where $\sigma_{s,t,g}$ is the standard deviation of all runs for this configuration.
To compute an aggregate score for each model across all configurations, we average the mean scores and derive the standard error of this aggregate score. Let $C$ be the set of all configurations, with cardinality $|C| = |S| \times |T| \times |G|$. The aggregate mean score $\bar{\mu}$ for a model is:
\begin{equation}
\bar{\mu} = \frac{1}{|C|} \sum_{(s,t,g) \in C} \mu_{s,t,g}
\end{equation}
For the standard error of this aggregate mean, assuming independence between configurations, we apply error propagation principles to obtain:
\begin{equation}
\text{SE}_{\bar{\mu}} = \sqrt{\frac{\sum_{(s,t,g) \in C} \text{SE}_{s,t,g}^{2}}{|C|^{2}}}
\end{equation}
This approach enables us to quantify both the average performance of each model across the entire benchmark and the statistical uncertainty associated with this estimate. We follow the recommendation of \citet{miller2024adding} and report the standard error of the mean for all our experimental results. In Figure \ref{fig:performance_by_cost}, we present the mean \benchmark{} score for each model with error bars indicating the standard error of the mean, allowing for comparison of model performance while accounting for statistical variability in the results.

\begin{figure}
    \centering
    \includegraphics[width=\linewidth]{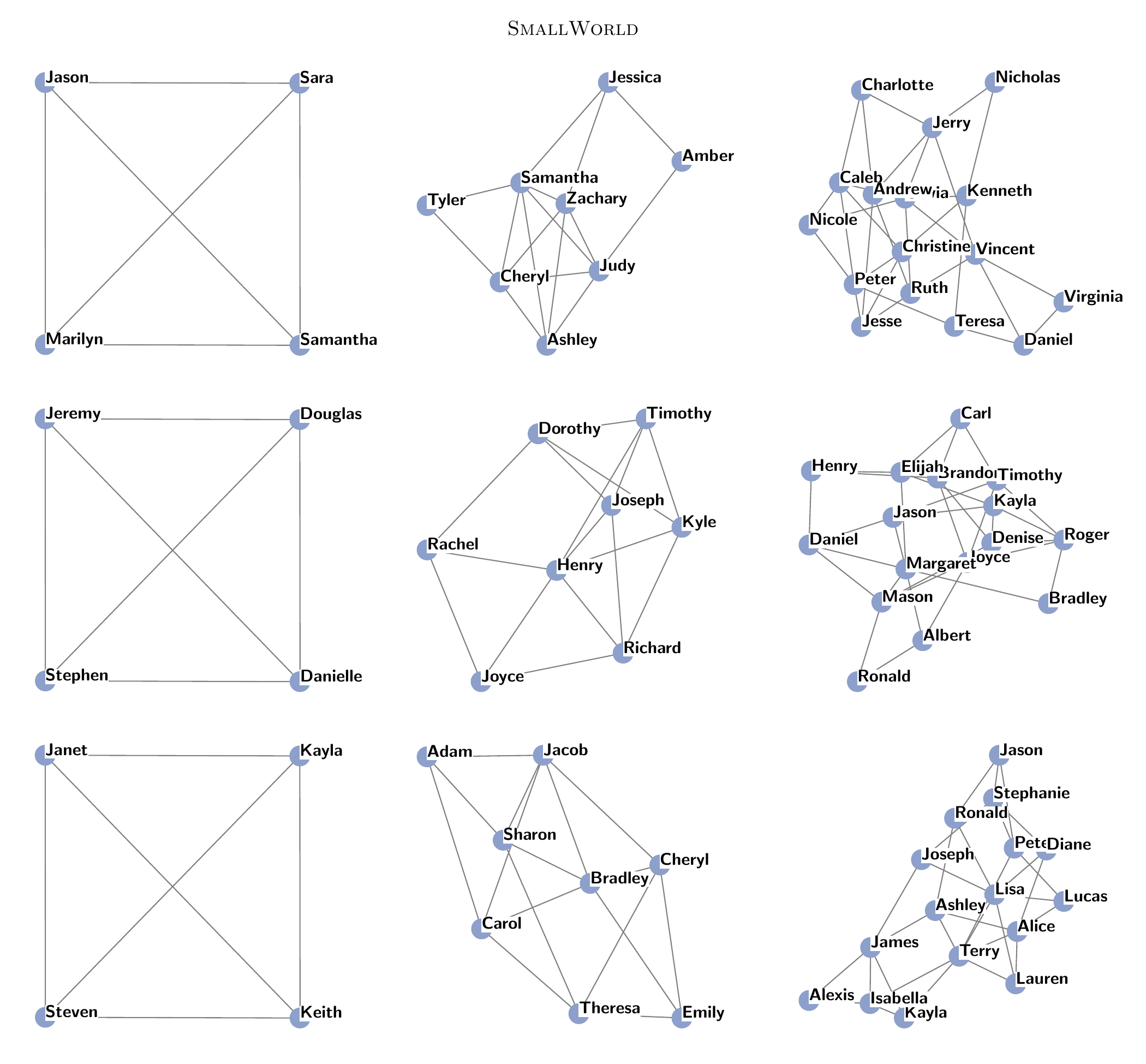}
    \caption{Network topologies of \benchmark{} generated from \textsc{SmallWorld} graphs.}
    \label{fig:ws_graphs}
\end{figure}

\begin{figure}
    \centering
    \includegraphics[width=\linewidth]{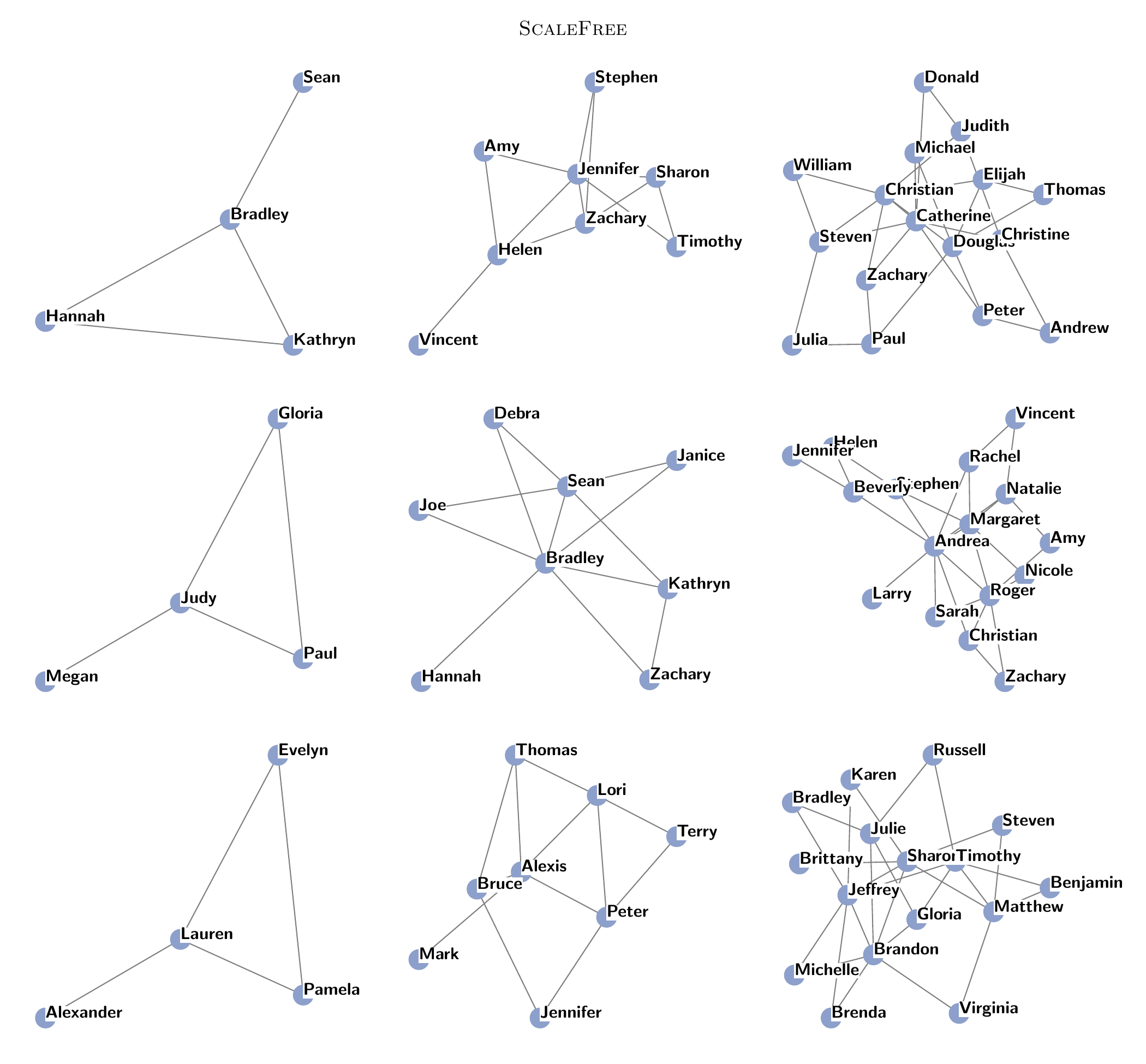}
    \caption{Network topologies of \benchmark{} generated from \textsc{ScaleFree} graphs.}
    \label{fig:ba_graphs}
\end{figure}

\begin{figure}
    \centering
    \includegraphics[width=\linewidth]{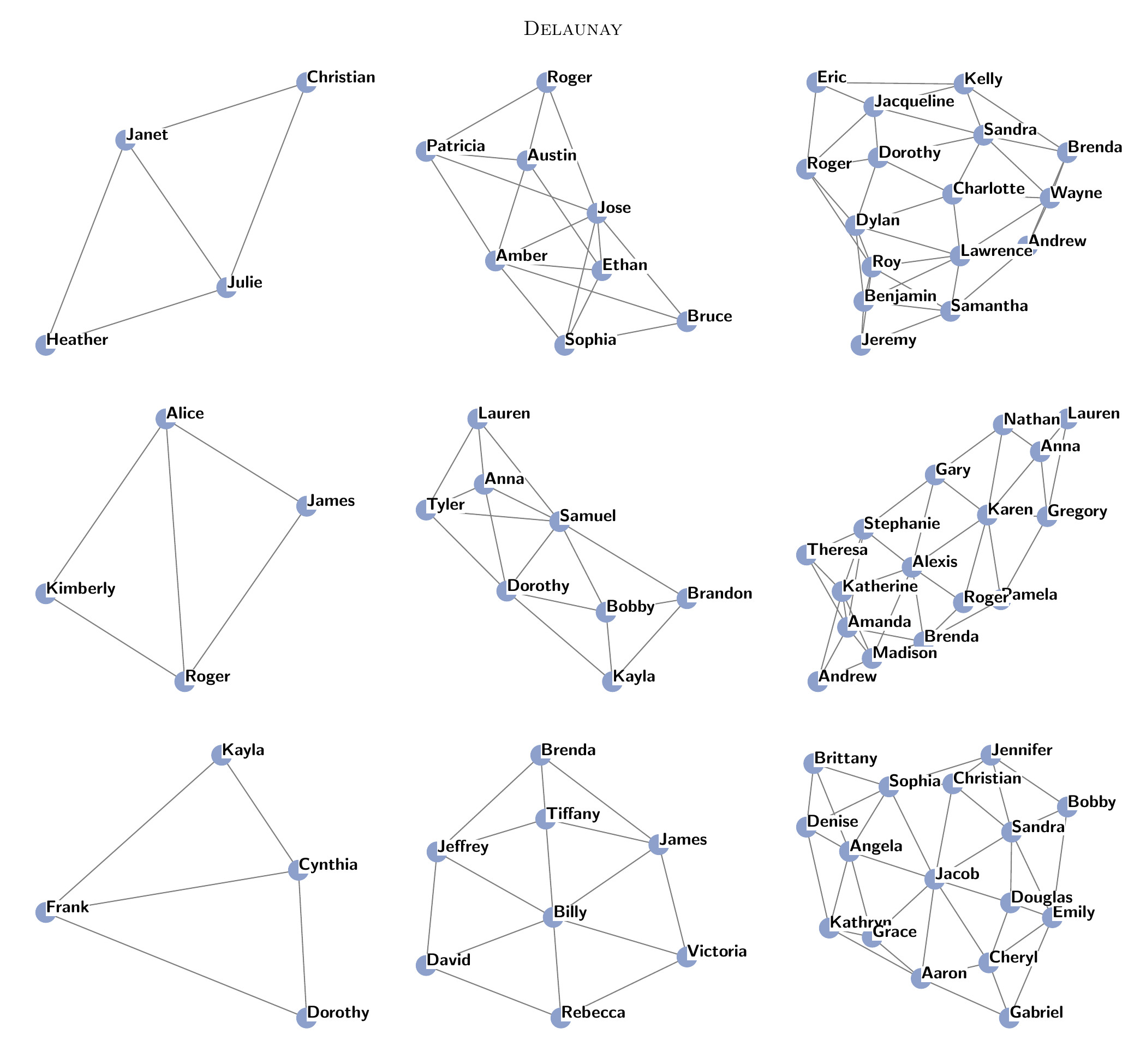}
    \caption{Network topologies of \benchmark{} generated from \textsc{Delaunay} graphs.}
    \label{fig:dt_graphs}
\end{figure}

\section{Graph Models}\label{app:graph_models}
Here, provide additional details about the graph models, as well as visualize the generated network topologies.

\paragraph{Small-world networks.} Generated using the Watts-Strogatz model~\citep{watts1998collective}, these graphs exhibit both short average path lengths and high clustering coefficients. They are commonly found in social networks, biological systems, and communication networks, making them highly relevant for studying agent-based interactions. In \benchmark{}, we refer to these graphs as \textsc{SmallWorld}; see \Cref{fig:ws_graphs} for a visualization of the network topologies generated from \textsc{SmallWorld} graphs.

\paragraph{Scale-free networks.} Constructed using preferential attachment models~\citep{barabasi1999emergence}, these graphs contain hubs (high-degree nodes) and follow a power-law degree distribution. Such networks are prevalent in the internet topology, citation networks, and organizational hierarchies, making them crucial for benchmarking coordination strategies.
Notably, \citet{kearns2006experimental} also draws a comparison between the two and claims that graphs generated with a preferential attachment model were more difficult to solve, although the graphs and testing they used were very limited. In \benchmark{}, we refer to these graphs as \textsc{ScaleFree}; see \Cref{fig:ba_graphs} for a visualization of the network topologies generated from \textsc{ScaleFree} graphs.

\paragraph{Geometric Graphs.}
We generate geometric graphs by constructing a Delaunay triangulation over randomly sampled 2D points. These graphs maintain a spatial relationship between nearby agents. Delaunay graphs are planar and connected, have relatively low maximum degrees, and short local distances, which makes them particularly relevant for modeling geographic networks, wireless communication, and distributed sensor systems. In contrast to small-world or scale-free graphs, geometric graphs emphasize proximity-based coordination, where collaboration is often limited by spatial locality rather than global topology. In \benchmark{}, we refer to these graphs as \textsc{Delaunay}; see \Cref{fig:dt_graphs} for a visualization of the network topologies generated from \textsc{Delaunay} graphs.

\section{Extended Qualitative Analysis}\label{app:qualitative}

\subsection{K4 or Star?}
Next, we present an illustrative example for a failure case arising from our key findings. Here, the task is \textsc{VertexCover} on a complete graph with 4 nodes and the model is o4-mini. After three rounds of message-passing, agents Douglas, Danielle, and Stephen have all determined the graph to be complete and agreed on the strategy to select three agents as the coordinators. However, in the last round of message-passing Jeremy sends messages to all neighbors, indicating the graph to be different:

\begin{chatlog}
\chatmsg[2]{Jeremy}{\texttt{\small The graph is actually a star: edges are only Jeremy–Douglas, Jeremy–Stephen, and Jeremy–Danielle. There are no leaf–leaf links. I therefore propose that I (Jeremy) be the sole coordinator and that you be a non-coordinator.}}
\end{chatlog}

This statement is directly contradictory to information Jeremy received in the last message-passing round. Moreover, based on this message, both Douglas and Stephen change their strategy as well as their belief about the structure and select Jeremy as the sole coordinator:

\begin{chatlog}
\chatmsg[2]{Douglas}{\texttt{\small Jeremy’s last message defines the graph
            as a star centered at Jeremy [...] If we choose only Jeremy as the
            sole coordinator, all edges are covered [...]}}
\end{chatlog}

Only Danielle recognizes Jeremy's suggested graph structure to be contradictory and sticks to the original strategy:

\begin{chatlog}
\chatmsg[2]{Danielle}{\texttt{\small Jeremy’s final claim that the graph is
            a star contradicts both Douglas and Stephen [...] Two independent agents (a
            majority) agree on K4, so that structure is more reliable [...]}}
\end{chatlog}

As a result, Douglas and Stephen each answer \textit{No}, and Danielle answers \textit{No} as Douglas, Stephen, and Jeremy were supposed to be coordinators under the original strategy. Only Jeremy answers \textit{Yes}. The resulting selection is not a vertex cover.

\subsection{Things Change}
We further present a negative example obtained with Gemini 2.5 Pro on the \textsc{Matching} problem for an 8-node graph.
The matching produced in this run is not maximal, as two neighboring agents, Kyle and Dorothy, both state that they did not find a neighbor to pair with.
Examining their conversation history, we observe that in round 2, the agents exchange the following messages to indicate that they both intend to match with other neighbors:

\begin{chatlog}
    \chatmsg[2]{Kyle}{\texttt{\small Hi Dorothy, thanks for your interest! I've just agreed to pair up with Timothy. Hope you find a partner soon!}}
    \chatmsg[2]{Dorothy}{\texttt{\small Hi Kyle, thanks for the message! I'm going to pursue a pairing with another neighbor who also reached out. Hope you find a partner soon!}}
\end{chatlog}

After this round, the communication between these two agents ceases, as both assume the other is no longer available.
However, both Kyle and Dorothy are unable to negotiate a pairing with other neighbors.
Despite this, no further messages are sent between the two.
In their final chains of thought, the agents explicitly base their final response on this outdated information:

\begin{chatlog}
    \chatmsg[2]{Kyle}{\texttt{\small [...] All my neighbors (Timothy, Henry, Dorothy, Richard) have either explicitly stated they paired with someone else or indicated they were pursuing other options [...]}}
    \chatmsg[2]{Dorothy}{\texttt{\small [...] Timothy, Kyle, and Rachel also confirmed they were paired with others. [...]}}
\end{chatlog}

This trace illustrates how agents appear to have persistent trust in the information received from neighbors, sometimes to a harmful degree.
Here, both agents did not consider that the situation of their neighbor may have evolved, just as their own.
As a consequence, they based their matching decisions on outdated information.

\subsection{Helping Others}
In the following, we provide an excerpt obtained from running Gemini 2.5 Pro on an 8-node graph coloring problem.
One agent, Kayla, appears to have noticed that two neighbors (Bobby and Brandon) do share an edge but plan to pick the same group.
Reacting to this, Kayla sends the following messages to each neighbor:

\begin{chatlog}
    \chatmsg[2]{Kayla (to Bobby)}{\texttt{\small Hi Bobby, got your
            message proposing Group 4. IMPORTANT: Brandon ALSO proposed
            Group 4 in his last message to me. Since you are neighbors,
            you CANNOT both be in Group 4. One of you MUST choose a
            different group for the final assignment. [...]}}
    \chatmsg[2]{Kayla (to Brandon)}{\texttt{\small Hi Brandon, got your message
            proposing Group 4. IMPORTANT: Bobby ALSO proposed Group 4 in
            his last message to me. Since you are neighbors, you CANNOT
            both be in Group 4. One of you MUST choose a different group
            for the final assignment. [...]}}
\end{chatlog}

This illustrates the apparent willingness of the agent to help others resolve their respective conflicts and achieve the shared goal of a conflict-free coloring. 
Note that Kayla was not prompted by any other agent to provide advice.
Instead, the agent self-initiated this helpful behavior.

\subsection{Sharing Challenges}
On several occasions, we observed agents explicitly discuss the challenges associated with the synchronous message passing setup, where all agents send messages simultaneously.
Here, we provide an example from Gemini 2.5 Pro on a graph coloring problem, where the agents, Samuel and Tyler, exchange the following comments about the message passing procedure:

\begin{chatlog}
    \chatmsg[2]{Samuel}{\texttt{\small Hi Tyler, thanks. Message lag is confusing. I proposed Group 3 last round reacting to the Group 2 pile-up. Now I see you, Lauren, Brandon, Bobby, and Anna are proposing Group 3, and Dorothy is proposing Group 1. [...]}}
    \chatmsg[2]{Tyler}{\texttt{\small Hi Samuel. Messages definitely crossed. My last message proposed Group 3, not Group 2, due to earlier conflicts. Now I see you, Lauren, and Anna are all proposing Group 3. [...]}}
\end{chatlog}

It appears that in this case, the agents recognized that simultaneous message exchange can cause synchronization to be challenging and that it may contribute to inconsistent information received from neighbors.
Furthermore, they exchange this thought with their neighbors on their own initiative, potentially alerting others to the issues they recognized.

\subsection{Priority 42}
With respect to strategy coordination, we observe that agents are trying to directly apply distributed computing algorithms. In the following example from o4-mini, the task is \textsc{Coloring} on 8 nodes. Here, the agent Henry decides to send the following first message to all neighbors:

\begin{chatlog}
    \chatmsg[2]{Henry}{\texttt{\small PRIORITY: 42. Please reply with your priority number.}}
\end{chatlog}

Note that Henry provides an elaborate explanation for this choice in the chain-of-thought:

\begin{chatlog}
    \chatmsg[2]{Henry}{\texttt{\small \textit{(CoT, not sent to neighbors)} Here’s my plan:  1. We need a way to break symmetry so that exactly one agent decides first, next, etc. A common approach is for each node to pick a random “priority” value and then to let higher‐priority nodes choose their colors before lower‐priority ones.   2. In Round 1 I’ll pick my own random priority value and broadcast it to all of my neighbors, and I’ll ask each of them to reply with their chosen priority.   3. Once I have everyone’s priorities (after Round 2), I’ll know my relative ordering among my neighbors. Later we can propagate these around the network so everyone learns enough to choose a valid color in the proper order.  So in this first round I send each neighbor my priority and ask for theirs ...}}
\end{chatlog}

Interestingly, despite the lack of explanation in Henry's message, neighboring agents are able to infer the proposed strategy:

\begin{chatlog}
    \chatmsg[2]{Kyle}{\texttt{\small \textit{(CoT, not sent to neighbors)} Henry is asking for my "priority number." In many distributed symmetry‐breaking schemes (e.g., for graph coloring), each agent picks a unique random priority. I can choose a number (say 7) to serve as my priority.}}
\end{chatlog}

Indeed, in the last round of message-passing Henry has gathered priority numbers from all neighbors:

\begin{chatlog}
    \chatmsg[2]{Henry}{\texttt{\small \textit{(CoT, not sent to neighbors)} [...] We know everyone’s priority (except Richard, who opted out, so we assume he’s not ahead of Henry in ordering).  The ordering by increasing priority is: Timothy(1) → Kyle(7) → Joseph(17) \& Rachel(17) → Henry(42) → Joyce(43) → Richard. [...]}}
\end{chatlog}

and describes the (implicit) strategy:

\begin{chatlog}
    \chatmsg[2]{Henry}{\texttt{\small \textit{(CoT, not sent to neighbors)} [...] Under the agreed greedy rule, each agent when it’s their turn picks the smallest group number not already used by its already‐colored neighbors. [...]}}
\end{chatlog}

First, it appears that the greedy rule has not indeed found consensus among the other agents. In particular, Timothy and Dorothy choose their group based on assumptions about neighboring group choices. 
Second, strategy coordination proves difficult in this example. While Kyle, Dorothy, Joseph, and Henry choose a group upfront and inform other agents about their choice, Joseph, Kyle, Timothy end up choosing a different group than they announced after hearing about other agents' group choices.

\section{Extended Related Work}
\label{appendix:extended_rw}
Recent research has increasingly focused on utilizing multiple LLM agents collaboratively to enhance performance and tackle complex problems.
``Multi-Agent Debate''~\citep{du2023improving,xiong2023examining,liang2024encouraging} allows multiple agents to iteratively discuss solutions, effectively acting as a parallelizable test-time computation scaling and self-consistency mechanism.
Further work introduces different network topologies for more structured agent interaction.
Some works study pre-determined graph structures \citep{regan2024problem,qian2024scaling}
while others propose to automatically adapt the network topology towards a given task \citep{liu2023dynamic,chen2024agentverse,zhuge2024gptswarm}.  
In particular, it has been observed that different network topologies work best for different tasks \citep{chen2024agentverse,zhuge2024gptswarm} and that, in some scenarios, the reasoning performance scales logistically in the network size \citep{qian2024scaling}.
The behavior of large-scale LLM agent networks has further been shown to resemble real social phenomena, such as misinformation spreading and herd effects \citep{yang2024oasisopenagentsocial,chuang2024simulating}.

Understanding the ability of LLMs to perform reasoning tasks on graph-structured data has become another active research area.
A range of studies propose datasets for evaluating LLMs on graph reasoning tasks \citep{fatemi2024talk,wang2024can,zhang2024llm4dyg,tang2025evaluating,skianis2024graphreasoninglargelanguage}.
These generally rely on a single-agent setup where a graph is encoded as text, and a single LLM instance is prompted to solve a particular reasoning task for this graph.
This setup is well-suited to study the capability of LLMs for solving complex tasks on structured data in a controlled setting.
\citet{fatemi2024talk} investigate the impact of how the input graph is encoded as text.
\citet{sanford2024understanding} categorize graph reasoning problems in terms of their depth- and width-complexity for transformer models.
\citet{wang2024can} and \citet{skianis2024graphreasoninglargelanguage} explore the effect of different prompting techniques for solving algorithmic graph problems.

Our work is positioned at the intersection of these two lines of research as we investigate how well multi-agent systems can collaboratively solve graph reasoning problems.
The consensus problem in multi-agent systems in a simple setting without text-based communication was studied by \citep{chen2023multi}.
Beyond this, the ability of multi-agent networks to collaboratively solve graph reasoning tasks has been investigated in \citet{xu2023magic} in the context of resource sharing. In contrast, \benchmark{} studies both coloring and vertex cover problems which can be instantiated as resource sharing tasks but additionally benefit from being theoretically well-studied and understood.
In addition, \benchmark{} is complementary to a range recent application-oriented agentic benchmarks~\citep{liu2024agentbench,yin2024mmau,agashe2024llm,yao2024tau,ni2025coral}

However, while those benchmarks focus on tasks involving mostly two agents, \benchmark{} is practically unlimited in size thanks to the generative protocol of problem creation and evaluation. 
Hence, \benchmark{} is harder to saturate as the size and complexity of problems can grow with the capabilities of frontier LLMs. 
For example, the current suite of problems involves 4, 8, and 16 agents but we also present experiments performed 
with 100 agents coordinating to solve a problem instance. 

In addition to a variety of benchmarks, there also exist multi-agent frameworks for LLMs, notably \citet{chen2024internet}, enabling LLMs to collaborate via a shared messaging platform, which supports, among other things, the formation of teams, a coordinative task similar to that of the matching problem we study in \benchmark{}. Further, \citet{chen2024agentverse} propose AgentVerse, demonstrating that collaborative multi-agent systems are able to outperform single agents.

Finally, a body of work exists investigating how human participants solve decentralized coordination problems in social networks. 
Experiments by \citet{kearns2006experimental} explore how human agents perform when tasked with negotiating a graph coloring and demonstrate a strong influence of the network topology on coordination success. 
\citet{judd2010behavioral} conducts similar studies for both graph coloring and the consensus problem and finds that the effect of the network topology on human performance is task-specific.
This line of studies was further extended to consider dynamically changing networks \cite{chiang2024adaptive}.

\section{Limitations}\label{app:limitations}
While \benchmark{} provides a principled and scalable benchmark for evaluating coordination and collaboration in multi-agent LLM systems, several limitations remain. 
The benchmark adopts a fixed and synchronous communication model based on the LOCAL framework, with all agents engaging in a pre-defined number of message-passing rounds. Although this choice aligns with theoretical work in distributed computing, it limits the ecological validity of the set-up. Many real-world multi-agent systems operate asynchronously or under dynamic communication constraints, and it remains unclear how well performance would transfer under such conditions.
Our evaluation protocol considers an instance solved only if it meets strict task-specific correctness criteria. This binary metric provides a clear signal for coordination success, but may obscure partial progress, particularly in tasks where near-correct solutions still demonstrate substantial reasoning capability. Moreover, while tasks are instantiated in diverse graph topologies, the agents themselves are homogeneous within each experiment, sharing architecture, capabilities, and prompting style. This homogeneity simplifies analysis, but does not capture heterogeneous agent settings, which are common in real-world deployments and pose additional coordination challenges.
Finally, the scalability of \benchmark{} is limited in practice by the computational cost of LLM inference. Although the benchmark can be instantiated with up to 100 agents, performance degrades significantly beyond small network sizes. This suggests that current LLMs are not yet capable of maintaining coherent global strategies under increasing communication and memory demands. In addition, the current setup assumes that all agents act cooperatively and faithfully follow the protocol. We do not consider settings with noisy, faulty, or adversarial agents, which would be essential for assessing robustness in more realistic deployments.

\section{Extended Results}\label{app:extended_results}

Table~\ref{tab:app_score_results} reports the soft scores per model and task. These scores capture partial correctness, offering a more granular view of model behavior than strict success/failure. However, soft scores are not directly comparable across tasks due to heterogeneous evaluation criteria and should only be interpreted within-task. Details on how these scores are computed are provided in Appendix~\ref{app:benchmark_tasks}.
Table~\ref{tab:main_results} presents the fraction of fully solved instances using the binary evaluation metric described in Section~\ref{sec:tasks_graphs_evaluation}. Compared to earlier results, Gemini 2.5 Pro shows consistently improved results and reaches a new state-of-the-art mean \benchmark{} score of 0.80. Although Gemini 2.5 Pro achieves a high average score, the results do not indicate saturation. In contrast, the small standard errors observed across the runs (Table~\ref{tab:main_results}) confirm that \benchmark{} remains well calibrated to distinguish between models of varying capabilities. Importantly, \benchmark{} is inherently scalable: By increasing the size of the graph, the benchmark can naturally be extended to match the capabilities of future models. This flexibility ensures that \benchmark{} can evolve alongside advances in multi-agent language systems and continue to provide meaningful performance differentiation.

\end{document}